\documentclass[10pt, journal]{IEEEtran}
\IEEEoverridecommandlockouts

\usepackage{amsmath,amssymb,amsfonts}
\usepackage{graphicx}
\usepackage{algorithmicx}
\usepackage{algorithm}
\usepackage{multirow}
\usepackage{threeparttable}  

\usepackage{algorithmicx}
\usepackage{algorithm}
\usepackage{algpseudocode}
\usepackage{changes}
\usepackage{anyfontsize}
\usepackage{array}
\usepackage[caption=false,font=normalsize,labelfont=sf,textfont=sf]{subfig}
\usepackage{textcomp}
\usepackage{stfloats}
\usepackage{url}
\usepackage{verbatim}
\usepackage{graphicx}
\usepackage{cite}
\usepackage{xcolor}
\usepackage{color}
\usepackage{array, makecell}
\usepackage{amssymb}
\usepackage{pifont}
\usepackage{booktabs}   
\usepackage{tabularx}   
\usepackage{float}      
\usepackage{caption}    
\usepackage{lipsum}     
\usepackage{booktabs}
\usepackage{pgf-pie}
\usepackage{caption}        
\usepackage{subcaption}     
\usepackage{subfig}
\usepackage{url}    
\usepackage{listings}
\usepackage{xcolor} 
\usepackage{balance}

\def\BibTeX{{\rm B\kern-.05em{\sc i\kern-.025em b}\kern-.08em
    T\kern-.1667em\lower.7ex\hbox{E}\kern-.125emX}}
    
\begin{document}
\setlength{\arrayrulewidth}{0.25pt}


\captionsetup[figure]{
  font=small,       
}
\captionsetup[table]{font=small}

\title{Understanding and Mitigating Errors of LLM-Generated RTL Code\\
\author{
    Jiazheng Zhang,
    Cheng Liu, \IEEEmembership{Senior Member, IEEE},
    Long Cheng, \IEEEmembership{Senior Member, IEEE}, \\
    Xiaowei Li, \IEEEmembership{Senior Member, IEEE}, 
    and Huawei Li, \IEEEmembership{Senior Member, IEEE}
    \thanks{
    
    Received 8 Augest 2025; revised 17 November 2025; accepted 28 January 2026. This work is in part supported by the Strategic Priority Research Program of the Chinese Academy of Sciences under Grant XDB0660103 and XDB0660100. (Corresponding author: Cheng Liu.)

    Jiazheng Zhang, Cheng Liu, Xiaowei Li and  Huawei Li are with the State 
Key Laboratory of Processors, Institute of Computing Technology, Chinese Academy of Sciences, Beijing 100190, China, and also with the Department of Computer Science and Technology, University of Chinese Academy of Sciences, Beijing 100190, China (e-mail: zhangjiazheng24@mails.ucas.ac.cn, liucheng@ict.ac.cn, lxw@ict.ac.cn, lihuawei@ict.ac.cn). Long Cheng is with the School of Control and Computer Engineering, North 
China Electric Power University, Beijing 102206, China.  (e-mail: lcheng@ncepu.edu.cn).

    Acknowledgements: Generative AI was used exclusively for language polishing in the preparation of this manuscript.
}
}
\date{}
\footnotetext[1]{* Corresponding author.}


}

\maketitle

\begin{abstract}
Despite the potential of large language model (LLM) based register-transfer-level (RTL) code generation, the overall success rate remains unsatisfactory, with limited understanding of the underlying causes of errors. 
To address this gap, we conduct a comprehensive analysis of errors in generated RTL code and manually categorize them based on their underlying causes.
Our findings reveal that most errors stem not from the reasoning capabilities of LLMs, but rather from a lack of RTL programming knowledge, insufficient understanding of circuit concepts, ambiguous design descriptions, or misinterpretation of complex multimodal inputs. Building on this insight, we leverage the in-context learning capabilities of foundational LLMs and propose a set of error correction techniques to address these issues effectively.
Specifically, we construct a domain-specific knowledge base and employ retrieval-augmented generation (RAG) to supply LLMs with the specialized knowledge required for RTL code generation. To mitigate errors arising from ambiguous descriptions, we introduce a set of design description rules and implement a rule-checking mechanism that helps clarify and complete the input specifications. For errors caused by misinterpretation of multimodal data, we integrate external tools that convert such inputs into meta-formats more compatible with LLM processing. For the remaining errors such as missing details of longer context, we adopt a classic iterative debugging loop i.e. simulation-error localization-error correction to gradually improve the generated RTL code. 
We incorporate these error correction techniques into a foundational LLM-based RTL code generation framework, resulting in significantly improved performance. Experimental results show that our enhanced framework achieves {98.1\%} accuracy on the VerilogEval benchmark {based on Deepseek-v3.2-Speciale}, demonstrating the effectiveness of our methods.
\end{abstract}


\section{Introduction}

\IEEEPARstart{R}{ecent} advances in Large Language Models (LLMs) \cite{llm} have significantly reshaped the landscape of automated code generation. Specialized code generation models such as Codex\cite{Codex}, CodeGen\cite{codegen} and foundational LLMs such as GPT\cite{gpt} and DeepSeek\cite{deepseek} have demonstrated impressive capabilities in producing syntactically correct and functionally meaningful code across a wide range of programming languages and tasks. However, while LLMs achieve high success rates for mainstream programming languages like Python, their performance on hardware description languages (HDLs) such as Verilog remains noticeably lower. This discrepancy is largely due to the scarcity of high-quality Verilog code in the pretraining corpora, as well as the inherent domain-specific complexity of RTL designs in general.

Nevertheless, errors in LLM-generated Verilog code arise from a variety of sources. Some stem from a lack of familiarity with Verilog-specific syntax and semantics\cite{syntax}. For example, the distinction between blocking and non-blocking assignments is critical in Verilog but largely absent in general-purpose programming languages. Other errors are rooted in the model’s limited understanding of circuit-level concepts. Concepts such as bit manipulation or sequential logic often require domain-specific expertise that LLMs, trained primarily on software code, may lack. For instance, operations based on bit series or timing-sensitive constructs can be difficult for LLMs to interpret and generate correctly. Given the diversity and complexity of these errors, a one-size-fits-all solution is unlikely to be effective. Therefore, a deeper understanding of the types and causes of these errors is essential for systematically improving the quality of LLM-based RTL code generation.

Recent research has taken important steps toward addressing these RTL code generation challenges. Approaches such as VeriGen \cite{Verigen} ,VerilogEval \cite{verilogeval}, RTLCoder \cite{rtlcoder}, {CodeV-R1}\cite{zhu2025qimeng} improve LLM performance by fine-tuning {and reinforcement learning} on larger domain-specific Verilog datasets. Other methods, including RTLFixer \cite{rtlfixer} and HDLDebugger \cite{hdldebugger}, adopt retrieval-augmented generation (RAG) strategies, incorporating external knowledge bases or repair examples to compensate for the models’ limitations in hardware design and debugging. Recognizing the difficulty of resolving errors in a single step, recent systems such as AutoChip \cite{autochip}, MEIC \cite{meic} {and MAGE}\cite{mage} employ iterative and multi-agent techniques to facilitate multi-round refinement and repair of LLM-generated Verilog code. Despite the varied methodologies, these approaches consistently demonstrate improved RTL code quality, suggesting that different techniques such as fine-tuning \cite{liu2024rtlcoder} and can effectively address similar errors. However, these advancements often lack a systematic analysis of the root causes of the errors, raising the question of whether such errors are truly resolved or merely happen to be corrected by chance. This gap hinders further improvement in LLM-based RTL code generation.

In this work, we present a comprehensive analysis of errors in LLM-generated RTL code, categorizing them manually according to their underlying causes. Our findings indicate that the majority of errors do not stem from the reasoning limitations of LLMs, but rather from a lack of RTL programming knowledge, inadequate ambiguous design specifications, misinterpretation of complex multimodal inputs or understanding of circuit design principles. Guided by these insights, we leverage the in-context learning capabilities of foundation models and introduce a set of targeted error correction techniques. 
Specifically, to address issues arising from ambiguous design descriptions, we define a set of specification rules and implement a rule-checking mechanism to ensure clarity and completeness. For errors caused by misinterpretation of multimodal data, we integrate external tools to convert such inputs into meta-formats better suited for LLM processing. Furthermore, we apply retrieval-augmented generation (RAG) to provide LLMs with the specialized RTL programming and circuit design knowledge required for precise interpretation of design specifications and accurate RTL code generation, supported by the construction of a domain-specific knowledge base.
Additionally, to handle residual errors such as those resulting from incomplete contextual understanding in case of longer context, we adopt a classic iterative debugging loop consisting of simulation, error localization, and correction. We integrate all these techniques into a foundational LLM-based RTL code generation framework, achieving substantial improvements in both code correctness and overall generation quality.

Our contributions are summarized as follows:

\begin{itemize}

\item We conduct an in-depth manual analysis of errors in LLM-generated RTL code and categorize them based on their root causes, revealing that most errors originate from insufficient domain knowledge rather than reasoning limitations of the models.

\item Based on the identified causes of errors, we propose a suite of correction techniques, including  a rule-based mechanism for description refinement, multimodal input conversion tools, a domain-specific knowledge base with RAG, and an iterative simulation-based debugging loop.

\item We integrate the error correction mechanisms into a representative LLM-based RTL code generation framework, achieving  {98.1\%} accuracy on the VerilogEval Benchmark based on {Deepseek-v3.2-Speciale} —— surpassing a series of RTL generation framework. 
\item We have open-sourced the erroneous RTL samples generated by LLMs, along with their labeled results, analysis reports, and corresponding error-correction code on GitHub \footnote{\url {https://github.com/zjz1222/RTLErrorAnalysis}}. We hope our work can contribute to advancing LLM-assisted hardware design.

\end{itemize}
\newcommand{\customsmall}{\fontsize{5.5}{7}\selectfont}
\lstdefinestyle{verilog-style}{
    language=Verilog,
    basicstyle= {\customsmall\ttfamily},
    keywordstyle=\color{blue},
    commentstyle=\color{black},
    stringstyle=\color{red},
    breaklines=true, 
    tabsize=4,
    showstringspaces=false,
    escapeinside=||
}

\section{Related Work}
\
\subsection{LLMs for RTL Code Generation}

With the rise of LLMs in the programming field, an increasing number of researchers are considering using LLMs for the automated generation of RTL designs from natural language descriptions. DAVE\cite{dave} emerged as an early study that fine-tuned GPT-2 \cite{gpt2} to automatically convert English specifications into Verilog code. Along this line, VeriGen\cite{Verigen} demonstrates that a fine-tuned open-source LLM (CodeGen-16B\cite{codegen}) can outperform GPT-3.5 Turbo \cite{gpt35} in generating functionally correct Verilog code, highlighting the potential of specialized smaller models for hardware design automation. Subsequently,
OriGen \cite{origen} enhances RTL dataset quality via code-to-code improvement. {CodeV-R1} \cite{zhu2025qimeng} {proposes a RLVR framework for training Verilog-generating LLMs.} {ReasoningV} \cite{qin2025reasoningv} {introduces a high-quality Verilog code dataset that incorporates reasoning.}
Following this, methods like Aivril\cite{aivril}, PromptV\cite{promptv}, and VerilogCoder\cite{verilogcoder} attempt to implement RTL generation using multi-agent approach. VRank\cite{vrank} and {MAGE}\cite{mage} also employ multi-candidate sampling method to assist high-quality RTL generation. HLSPilot\cite{xiong2024hlspilot} also introduced an LLM-based framework for HLS generation.
\vspace{-1em}
\subsection{LLMs for RTL Code Debugging}
Despite the promising RTL code generation capabilities of LLMs, the success rate of the LLM-based RTL code generation remains unsatisfactory. An intuitive approach is to repeat the code generation procedures multiple times and select the best for the code generation, but this approach generally poses limited improvement and there is no guarantee for better results due to the uncertainty of LLM-based generation. Hence, debugging becomes a critical procedure to the LLM-based RTL code generation, and tremendous efforts have been devoted to fix the bugs and enhance the generated results continuously. For example, RTLFixer \cite{rtlfixer} combines the ReAct framework and RAG technology to fix compilation errors using a knowledge base of expert insights. HDLDebugger \cite{hdldebugger} enhances LLM error correction by integrating RAG with fine-tuning and a code sample library. AutoChip \cite{autochip} employs a multi-round iterative approach guided by compilation and simulation feedback, while MEIC\cite{meic} employs a multi-agent framework with iterative refinement to correct both syntactic and functional errors in Verilog code. 
VeriDebug\cite{wang2025veridebug} introduces a contrastive representation and guided correction approach for automated Verilog debugging.

Prior studies have demonstrated the significant potential of using LLMs for RTL code generation and debugging. However, LLMs remain susceptible to introducing errors during RTL generation, and the underlying reasons of these errors remain insufficiently understood. Moreover, there is a lack of systematic characterization of the causes and manifestations of errors introduced during RTL generation with foundational LLMs. Such characterization is crucial for identifying major challenges and guiding future research on LLM-based RTL code generation. In the next section, we will explore an important question:  \textit{Why do LLMs produce errors in RTL generation?}

\section{Error Analysis of LLM-Generated RTL Code}


\begin{table}[htbp]
    \centering
    \begin{minipage}{0.48\textwidth}
        \centering
        \caption{{Statistics of faulty design instances for different models on the VerilogEval benchmark}}
        \label{tab:model_errors}
        \begin{tabular}{ >{\centering\arraybackslash}m{1.4cm}| >{\centering\arraybackslash}m{3.9cm}| >{\centering\arraybackslash}m{1.0cm}| >{\centering\arraybackslash}m{1.0cm} }
            \toprule
            \textbf{Category} & \textbf{Model Name} & \textbf{Release Date}\footnotemark[1] & \textbf{Faulty Design} \\
            \midrule
            \multirow{3}{*}{\textit{General}}
            & DeepSeek-V3.2-Speciale\cite{liu2025deepseek} & 2025.12 & 17 \\
            & GPT-4 Turbo\cite{gpt4} & 2024.04 & 65 \\
            & GPT-3.5 Turbo\cite{gpt35} & 2024.01 & 89 \\
            \midrule
            \multirow{1}{*}{\textit{Code-Specific}}
            & Qwen2.5-Coder-32B-Instruct\cite{hui2024qwen2} & 2024.11 & 83 \\
            \midrule
            \multirow{1}{*}{\textit{RTL-Specific}}
            & CodeV-R1-RL-Qwen-7B\cite{zhu2025qimeng} & 2025.06 & 52 \\
            \bottomrule
        \end{tabular}
        \vspace{0.5em}
        \vspace{-1em}
        \footnotesize
        \footnotetext[1]{The release date corresponds to the latest version of the model available as of the study.}
    \end{minipage}
\vspace{-1em}
\end{table}

To understand the RTL code generation capabilities of foundational LLMs, we analyze the errors in the RTL code produced by these models and investigate their root causes. 



{Based on the VerilogEval benchmark, we conduct a comprehensive evaluation of the current large language models' capability to generate RTL code by employing a diverse set of models that vary in training dataset, parameter scale, and reasoning ability. The evaluation results are presented in Table \ref{tab:model_errors}.}

{Our assessment strategy is tailored to the characteristics of the models. For standard text-completion models without reinforcement learning fine-tuning for reasoning tasks, such as \textit{GPT-3.5 Turbo}, \textit{GPT-4 Turbo}, and \textit{Qwen2.5-Coder-32B-Instruct}, we adopt a direct Natural Language to Code (NL2Code) approach. For models specifically trained with reinforcement learning for reasoning tasks, like \textit{DeepSeek-V3.2-Speciale} and \textit{CodeV-R1-RL-Qwen-7B}, we utilize their recommended Reasoning configurations to fully leverage their code generation potential. The generated RTL code is verified and simulated using \textit{Icarus Verilog}.}

{From the test results, a total of 306 faulty design instances are collected and analyzed.} We manually examine the failed cases and classify them into two primary categories as listed below.

\definecolor{lightred}{rgb}{1, 0.88, 0.89}
\definecolor{lightgreen}{rgb}{0.88, 1, 0.89}

\begin{table*}[h!] 
\small
\centering
\vspace{-1em}
\caption{{Analysis of errors induced by insufficient knowledge of specialized RTL programming}}
\label{tab:error_stats}
\footnotesize
\renewcommand{\arraystretch}{0.1} 
\setlength{\tabcolsep}{0.5em} 
\begin{tabular}{
>{\centering\arraybackslash}m{2.6cm}|
>{\centering\arraybackslash}m{5.0cm}|
>{\centering\arraybackslash}m{4.0cm}|
>{\centering\arraybackslash}m{0.8cm}|
>{\centering\arraybackslash}m{1.0cm}|
>{\centering\arraybackslash}m{0.8cm}|
>{\centering\arraybackslash}m{0.8cm}|
>{\centering\arraybackslash}m{0.8cm}
}

\toprule
\textbf{Error Category} & \textbf{Error Examples} & \textbf{Golden Generation} &
\textbf{{DS}} & \textbf{GPT-4} &\textbf{GPT-3.5} &\textbf{Qwen} & \textbf{{CodeV-R1}}
\\
\midrule
Wire in Always Block
&\begin{minipage}[htbp]{0.4\textwidth} 
\vspace{-0.8em}
\begin{lstlisting}[style=verilog-style]
output  out;
always @(*) begin ...
    case (state) ...
    B: |\colorbox{lightred}{out = 1;}|
end
\end{lstlisting}
\vspace{-0.8em}
\end{minipage} 
&
\begin{minipage}[htbp]{0.4\textwidth} 
\vspace{-0.8em}
\begin{lstlisting}[style=verilog-style]
output out;
assign |\colorbox{lightgreen}{out = (state == B);}| 
\end{lstlisting}
\vspace{-0.8em}
\end{minipage} & 0 & 1 & 8 & 9 & 0
 \\
\midrule[0.25pt]
Numerical Processing Logic Error
&
\begin{minipage}[htbp]{0.4\textwidth} 
\vspace{-0.8em}
\begin{lstlisting}[style=verilog-style]
// Count the number of 1s in the first ...
assign |\colorbox{lightred}{count1 = in[1:0];}|
// Count the number of 1s in the last ...
assign  count2 = count1 + in[2];
assign  out = count2;
\end{lstlisting}
\vspace{-0.8em}
\end{minipage} 
&
\begin{minipage}[htbp]{0.4\textwidth} 
\vspace{-0.8em}
\begin{lstlisting}[style=verilog-style]
assign |\colorbox{lightgreen}{out = in[0]+in[1]+in[2];}|
\end{lstlisting}
\vspace{-0.8em}
\end{minipage} 
& 0 & 0 & 1 & 2 & 1
\\
\midrule[0.25pt]
Vector Bit Selection Error
&
\begin{minipage}[htbp]{0.4\textwidth} 
\vspace{-0.8em}
\begin{lstlisting}[style=verilog-style]
module TopModule (input clk, 
...  output [3:1] g
);
     assign  g[1] = (state == B);
     assign |\colorbox{lightred}{g[0] = 1'b0;}|
\end{lstlisting}
\vspace{-0.8em}
\end{minipage} 
&
\begin{minipage}[htbp]{0.4\textwidth} 
\vspace{-0.8em}
\begin{lstlisting}[style=verilog-style]
module TopModule (input clk, 
...  output [3:1] g
);
     assign g[1] = (state == B);
\end{lstlisting}
\vspace{-0.8em}
\end{minipage} 
& 0 & 0 & 3 & 1 & 1
\\
\midrule[0.25pt]
Inversion of Vector Slice Selection
&
\begin{minipage}[htbp]{0.4\textwidth} 
\vspace{-0.8em}
\begin{lstlisting}[style=verilog-style]
module TopModule (
    input [99:0] in, ...
    output [99:1] out_any,
);
    assign |\colorbox{lightred}{out\_any = in[99:1] \textbar\ in[0:98];}|
\end{lstlisting}
\vspace{-0.8em}
\end{minipage} 
& 
\begin{minipage}[htbp]{0.4\textwidth} 
\vspace{-0.8em}
\begin{lstlisting}[style=verilog-style]
module TopModule (
  input [99:0] in, ...
  output [99:1] out_any,
);
  assign |\colorbox{lightgreen}{out\_any = in[99:1] \textbar\ in;}|

\end{lstlisting}
\vspace{-0.8em}
\end{minipage} 
& 1 & 0 & 1 & 1 & 0
\\

\midrule[0.25pt]
Incomplete RTL Code
&\begin{minipage}[htbp]{0.4\textwidth} 
\vspace{-0.8em}
\begin{lstlisting}[style=verilog-style]
assign out =  |\colorbox{lightred}{(sel == 0) ? in[0]:}|
              |\colorbox{lightred}{(sel == 1) ? in[1]:}|
//Continue this pattern for all 256 inputs
              |\colorbox{lightred}{in[255]};|
\end{lstlisting}
\vspace{-0.8em}
\end{minipage} 
&\begin{minipage}{0.4\textwidth} 
\vspace{-0.8em}
\begin{lstlisting}[style=verilog-style]
assign out = |\colorbox{lightgreen}{in[sel]}|;
\end{lstlisting}
\vspace{-0.8em}
\end{minipage} &
 0 & 3 & 4 & 0 & 0
\\
\midrule[0.25pt]
Variable Redefinition
&\begin{minipage}[htbp]{0.4\textwidth} 
\vspace{-0.8em}
\begin{lstlisting}[style=verilog-style]
module TopModule(input clk,
...,   output reg [7:0] q);
|\colorbox{lightred}{reg [7:0] q;}|
\end{lstlisting}
\vspace{-0.8em}
\end{minipage} 
& \begin{minipage}[htbp]{0.4\textwidth} 
\vspace{-0.8em}
\begin{lstlisting}[style=verilog-style]
module TopModule(input clk,
...,   output reg [7:0] q);
\end{lstlisting}
\vspace{-0.8em}
\end{minipage} 
& 0 & 0 & 4 & 0 & 0
\\
\midrule[0.25pt]
Using Undefined Variables
&\begin{minipage}[htbp]{0.4\textwidth} 
\vspace{-0.8em}
\begin{lstlisting}[style=verilog-style]
module TopModule (input c, d,
    output [3:0] mux_in
);  
    wire w1, w2 ...;
    assign mux_in[0] = |\colorbox{lightred}{a \& b}| ? w1 : w2;
//The variables a and b are undefined.
\end{lstlisting}
\vspace{-0.8em}
\end{minipage} 
& \begin{minipage}[htbp]{0.4\textwidth} 
\vspace{-0.8em}
\begin{lstlisting}[style=verilog-style]
module TopModule (input c, d,
    output [3:0] mux_in
);
    assign mux_in[0] =  |\colorbox{lightgreen}{c \textbar\ d}|;
\end{lstlisting}
\vspace{-0.8em}
\end{minipage} 
& 0 & 1 & 1 & 0 & 0
\\
\midrule[0.25pt]
Incorrect Use of Generate Statement
&\begin{minipage}[htbp]{0.4\textwidth} 
\vspace{-0.8em}
\begin{lstlisting}[style=verilog-style]
module TopModule(input clk, ...,
    output reg [511:0] q);
always @(posedge clk) begin ...
   |\colorbox{lightred}{genvar i;}|
   |\colorbox{lightred}{generate}|
    for (i = 1; i < 511; i = i + 1)
   |\colorbox{lightred}{begin: gen\_rule90}|
   |\colorbox{lightred}{always @(posedge clk) }|
        q[i] <= q[i-1] ^ q[i+1];
   |\colorbox{lightred}{end}|
   |\colorbox{lightred}{endgenerate}| ...
end
\end{lstlisting}
\vspace{-0.8em}
\end{minipage} 
& \begin{minipage}[htbp]{0.4\textwidth} 
\vspace{-0.8em}
\begin{lstlisting}[style=verilog-style]
module TopModule(input clk, ...,
    output reg [511:0] q);
always @(posedge clk) begin
   |\colorbox{lightgreen}{integer i;}|
    for (i = 1; i < 511; i = i + 1)
    begin
        q[i] <= (q[i-1] ^ q[i+1]);
    end
end
\end{lstlisting}
\vspace{-0.8em}
\end{minipage} 
& 1 & 1 & 1 & 0 & 1
\\
\\
\bottomrule
\end{tabular}
\vspace{-2em}
\end{table*}

\begin{itemize}
\item \textbf{Type I: Insufficient Knowledge of Specialized RTL Programming (IKSP.):}
In these cases, the LLM correctly interprets the design specification but fails due to a lack of specialized knowledge in RTL programming. If the same specification were used to generate code in a more widely used language like Python, the generation would likely succeed.

\item \textbf{Type II: Misinterpretation of Design Specifications (MDS.):}
Here, the LLM fails to correctly interpret the design intent, which may stem from either insufficient understanding of circuit design concepts or general miscomprehension of the specifications. This type of error would likely persist even if a high-level language like Python were used.

\end{itemize}

\begin{figure*}[htbp]
\vspace{-1em}
\centering
\includegraphics[width=1.0\textwidth]
{./img/pie.png}
\vspace{-1.5em}
\caption{{The distribution of error type ratios in RTL code generation scenarios.}}
\label{fig:pie}
\vspace{-1.5em}
\end{figure*}

For different categories of LLMs, we conducted a detailed statistical analysis of the distribution of the different aforementioned error types, as shown in Fig \ref{fig:pie}. {Increasing model scale or domain-specific RTL training can effectively reduce Type I errors. However, Type II errors — particularly insufficient understanding of circuit concepts — remain the primary bottleneck limiting LLMs' ability to correctly generate RTL code.}


\vspace{-1em}
\subsection{Insufficient Knowledge of Specialized RTL Programming}
Although Verilog shares similarities with C-like languages, it incorporates specialized syntax and is deeply intertwined with circuit design concepts, making it substantially different from conventional high-level languages such as Python or C/C++. For example, Verilog enforces distinct assignment semantics for different contexts: blocking assignments and non-blocking assignments are used for combinational and sequential logic, respectively. Foundational LLMs often struggle with such distinctions due to the limited availability of Verilog-specific data in their training corpora. As a result, many RTL coding errors stem from insufficient understanding of RTL programming. According to our experiments on VerilogEval, we identified a total of 41 failed cases, which can be roughly categorized into four major groups, as shown in Table \ref{tab:error_stats}. 

\textbf{Wire in Always Block:}
The LLM often fails to correctly interpret RTL variable usage specifications, leading to incorrect assignments to variables declared as wire types within always blocks, as shown in Table~\ref{tab:error_stats}. This type of error is more likely to occur in smaller-scale models such as Qwen-Coder-32B-Instruction. However, for larger-scale models like GPT-4 Turbo, as the training corpus and model parameters increase, the frequency of such errors decreases significantly.


\textbf{Numerical Processing Logic Error:}
In some design contexts, LLMs are prone to errors in numerical processing logic. For instance, as illustrated in the Table ~\ref{tab:error_stats}. When implementing the function of counting the number of 1's in the vector in[1:0], the LLM incorrectly equates in[1:0] with in[1] + in[0]. 

\textbf{Vector Bit Selection Error:}
In some design scenarios, LLMs are prone to boundary violations when selecting bits from vectors.

\textbf{Inversion of Vector Slice Selection:}
Likewise, in some RTL design contexts, LLMs occasionally perform vector slicing in a direction inconsistent with the original vector definition, leading to compilation failures.

\textbf{Incomplete RTL Code:}
The LLM correctly interprets the design specifications, but generates only a portion of the required code. It often omits the remaining code inappropriately, particularly when the expected output is lengthy and repetitive. As shown in Table \ref{tab:error_stats}, the model uses a comment to substitute the remaining branching cases in a typical switch block, instead of generating all 256 possible branches.

\textbf{Variable Redefinition:}
LLM-generated RTL code often exhibits redefinitions of variables, particularly when the same variables serve as both registers and output signals. This type of error occurs relatively frequently in GPT-3.5 Turbo.

\textbf{Using Undefined Variables:}
In some RTL designs, LLMs may improperly reference undeclared variables during RTL generation. As shown in Table \ref{tab:error_stats}, when implementing the mux\_in[0] signal, the LLM incorrectly uses undefined variables 'a' and 'b', resulting in compilation errors.

\textbf{Incorrect Use of Generate Statement:}
In some cases, LLMs may confuse 'generate for' constructs with regular 'for' statements. The case in Table \ref{tab:error_stats} shows erroneous usage where the LLM attempted to iteratively update individual bits of register 'q' through generate statements, violating synthesis rules and leading to malfunction.




{Based on the statistics presented above, distinct error patterns are identified across the models: \textit{Qwen2.5-Coder-32B-Instruct} is prone to flawed logic in numerical and vector operations; \textit{GPT-4 Turbo} tends to produce incomplete code or misuse advanced RTL syntax such as \texttt{generate} blocks; \textit{GPT-3.5 Turbo} exhibits a wide variety of errors; and \textit{CodeV-R1-RL-Qwen-7B} shows susceptibility to mistakes in specific numerical and vector processing designs.}

{We also observe that both large models fine-tuned with RL reasoning for programming tasks and smaller models trained specifically in the RTL domain contribute to mitigating the issue of Insufficient Knowledge of Specialized RTL Programming.}



\vspace{-1em}
\subsection{Misinterpretation of Design Specifications}
Beyond the errors resulting from insufficient knowledge of specialized RTL programming, the majority of issues arise from misinterpretation of design specifications. These misinterpretations can be categorized into four main types.
Detailed error categories will be illustrated in the rest of the subsections.



\subsubsection{Ambiguous Design Descriptions (ADD.)}

\begin{figure}[htbp]
\centering
\includegraphics[width=1.0\columnwidth]
{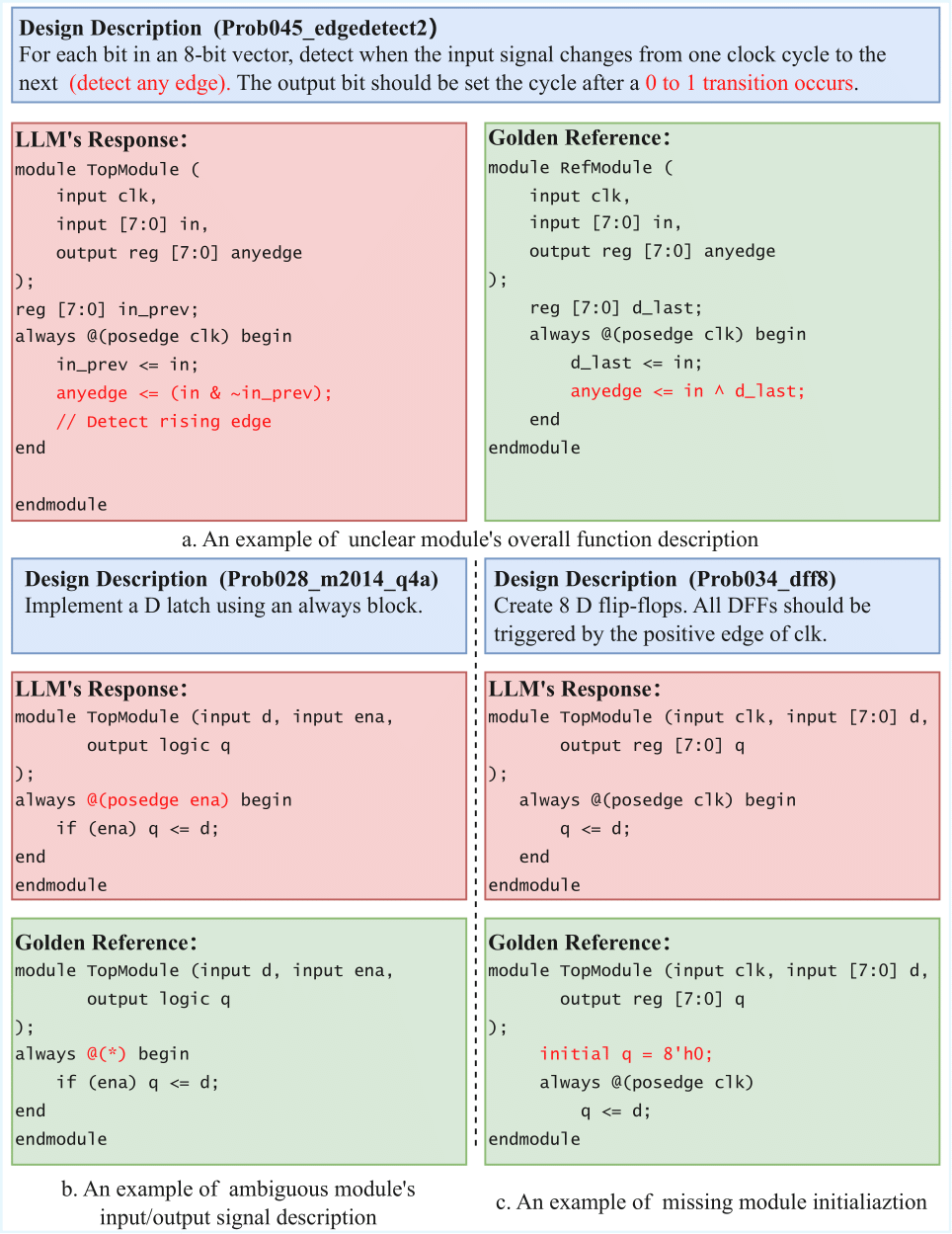}
\vspace{-1.5em}
\caption{Examples of ambigious design descriptions}
\vspace{-2em}
\label{fig:ami}
\end{figure}

We observe that a significant portion of errors stem from ambiguous design descriptions. In such cases, the LLM may interpret the ambiguous information differently from the fixed golden reference implementation. 
Ambigious design descriptions can be primarily categorized into three aspects: unclear overall module functionality, ambiguous module input/output signal descriptions, and lack of initialization.

\textbf{Unclear Overall Module Functionality (UOMF.): } 
This type of error primarily manifests as internal contradictions in the overall functional description of the module or inconsistencies between the description and the Golden Design implementation. As shown in Fig. \ref{fig:ami}(a), the first part of the design description requires detecting any edge, while the end specifies capturing only rising edges (0 to 1). {In the benchmark, the golden design implements "detect any edge", while the LLM implements "0 to 1 transition occurs". The ambiguity in the overall functional description within natural language leads to the LLM-generated design being evaluated as incorrect during evaluation.}


\textbf{Ambiguous Module Input/Output Description (AIOD.):}
This type of error manifests as insufficient description of the basic functionality of each input/output signal. As shown in Fig. \ref{fig:ami}(b), the design description does not clearly specify the triggering mechanism of ena. {In the benchmark, the golden implementation is edge-triggered, while the LLM-generated design is level-triggered. This discrepancy leads to the LLM's output being judged as a functional error during evaluation.}


\textbf{Missing Module Initialization Configuration (MMI.): }
Some design errors are caused by the description omitting module initialization information. As shown in Fig. \ref{fig:ami}(c), {since the description did not specify initialization requirements, the LLM implemented the design strictly according to the given description. However, due to extra initialization in the golden design, the LLM-generated design was incorrectly judged as functionally incorrect during evaluation.}


\begin{figure}[htbp]
\vspace{-0.8em}
\centering
\includegraphics[width=0.9\columnwidth]
{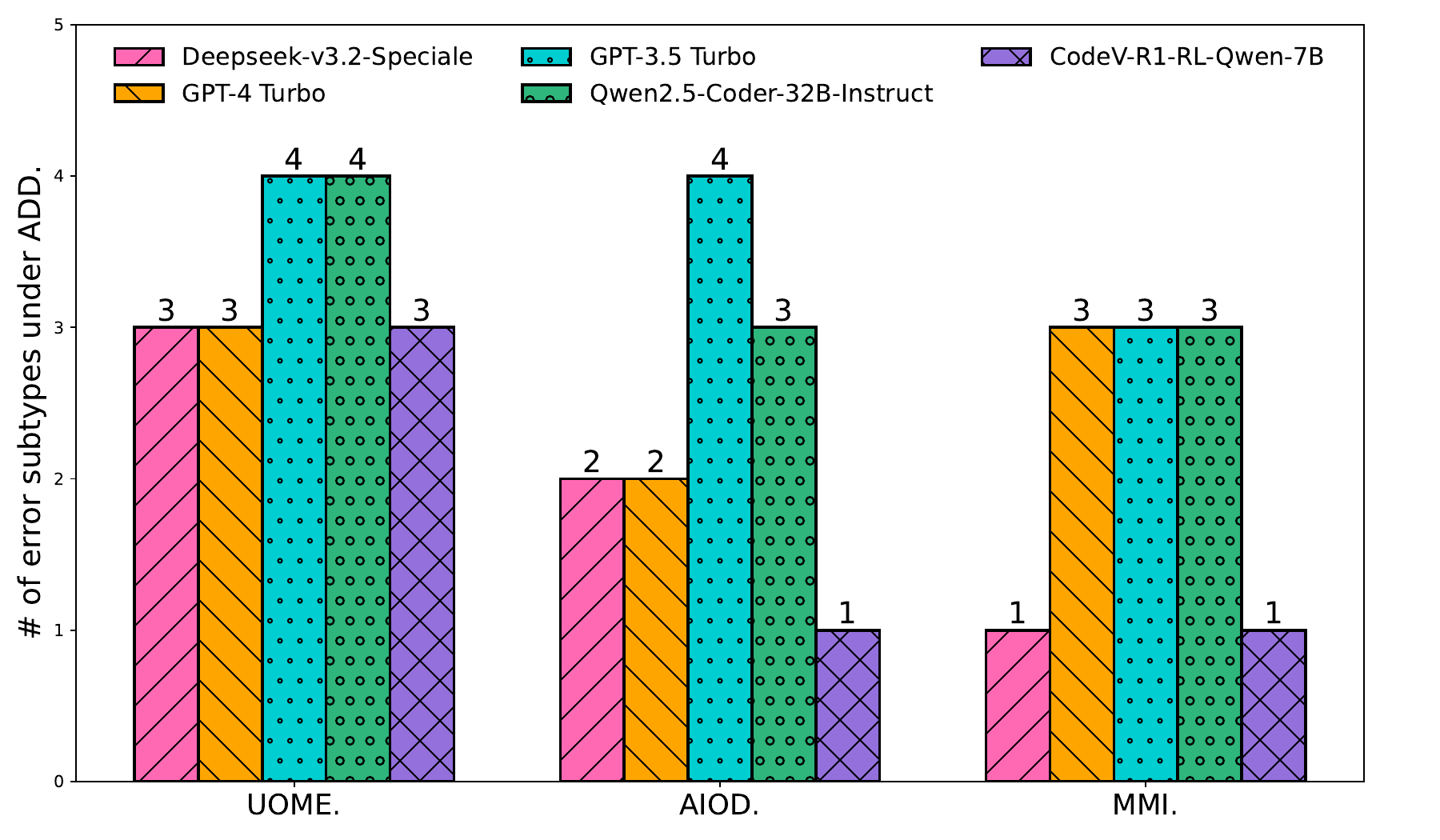}
\vspace{-1em}
\caption{{Distribution of error subtypes under ADD.}}
\vspace{-1em}
\label{fig:amic}
\end{figure}

We analyzed the distribution of subcategory errors under ambiguous design descriptions, as shown in Fig \ref{fig:amic}. The results indicate that, regardless of model capability, all models exhibit similar performance across different error categories. 


\begin{figure}[htbp]
\vspace{-1em}
\centering
\includegraphics[width=0.9\columnwidth]
{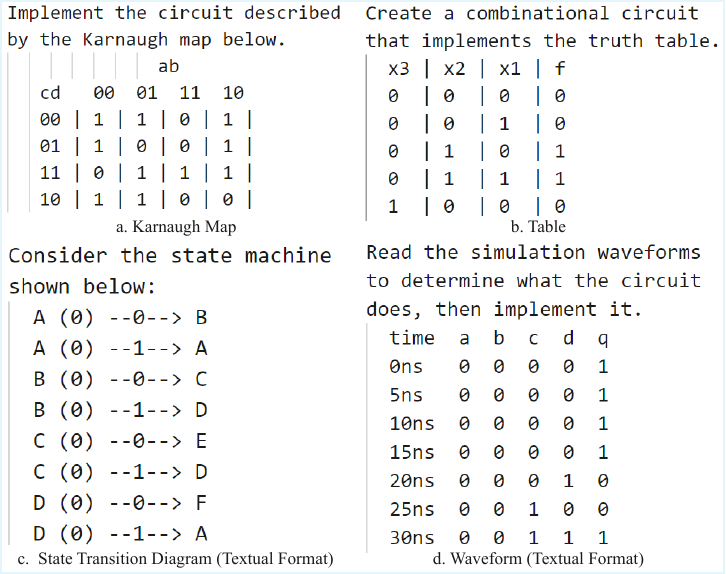}
\caption{Examples of multimodal data in hardware circuit design}
\vspace{-0.5em}

\label{fig:multimodal}
\end{figure}

\subsubsection{Misinterpretation of Multimodal Data (MMD.)}
Another significant source of code generation errors arises from the presence of multimodal data in design descriptions, such as \textbf{Karnaugh maps (KMAP.)},  \textbf{Tables (TB.)}, \textbf{State Transition Diagrams (ST.)}, and \textbf{Waveform Diagrams (WAV.)}, as illustrated in Fig. \ref{fig:multimodal}.
We observed that when LLMs process data in the aforementioned format, they struggle to accurately convert such multimodal information into correct RTL code due to limitations in their training data distribution.

\begin{figure}[htbp]
\vspace{-0.8em}
\centering
\includegraphics[width=0.9\columnwidth]
{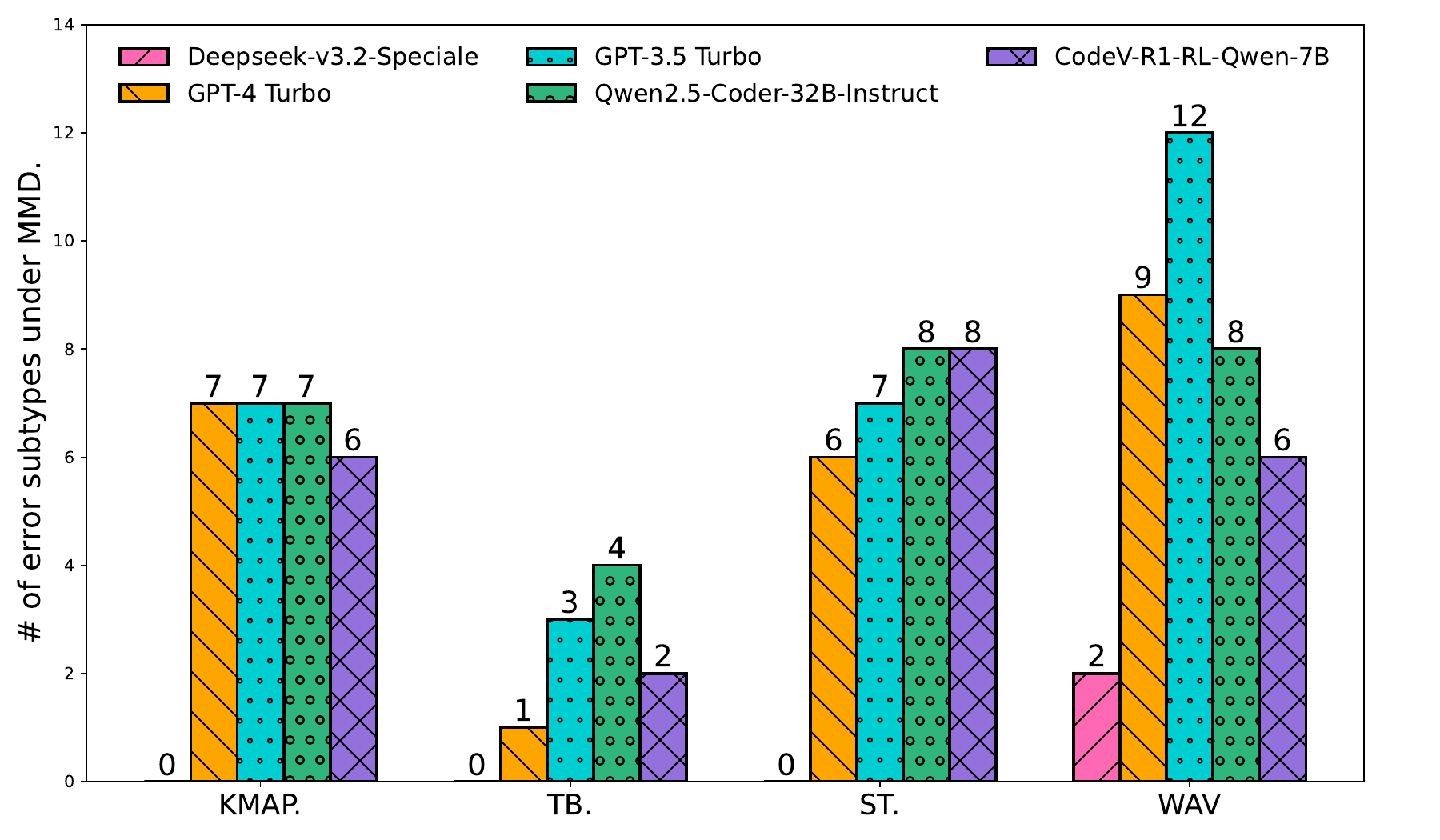}
\vspace{-1.0em}
\caption{{Distribution of error subtypes under MMD.}}
\vspace{-1em}
\label{fig:amic}
\end{figure}

We analyzed the distribution of subcategory errors caused by misinterpretation of multimodal information. The results show that all LLMs {except \textit{Deepseek-v3.2-Speciale}} performed poorly in comprehending multimodal information, with each model exhibiting over 20 design errors attributable to such misinterpretations. {Due to its excellent comprehension and reasoning capabilities, \textit{DeepSeek-V3.2-Speciale} generated only two erroneous designs when provided with Waveform format inputs.}

\subsubsection{Insufficient Understanding of Circuit Concepts (IUCC.)}
In this case, the LLM fails to correctly interpret the specification due to the absence of key circuit design concepts in the description, leading to incorrect code generation. Furthermore, the lack of circuit-related concepts can be categorized into four main causes: inadequate grasp of timing principles, insufficient understanding of specific design, limitations in numerical and vector processing capabilities, and deficiencies in state machine design for complex scenarios. These gaps collectively lead to constrained comprehension of design specifications.

\begin{figure}[htbp]
\vspace{-1em}
\centering
\includegraphics[width=1.0\columnwidth]
{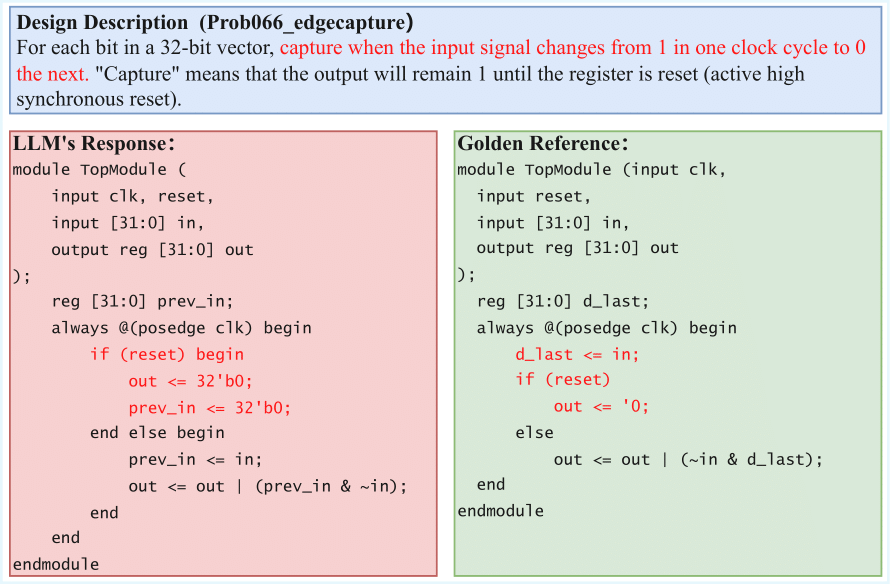}
\vspace{-1em}
\caption{Examples of inadequate understanding of timing related concept}
\vspace{-1em}
\label{fig:trc}
\end{figure}

\textbf{Inadequate Understanding of Timing Related Concepts (TRC.): }
In this scenario, the LLM's lack of domain-specific knowledge in timing-related concepts leads to functional errors in the generated RTL code. These errors include mishandling synchronous/asynchronous logic, improper previous signal latching, and incorrect timing distinctions in assign vs. always block assignments. 
As illustrated in Fig. \ref{fig:trc}, because the LLM lacks knowledge about tracking the predecessor signal (prev\_in) in edge detection, it incorrectly clears prev\_in within the reset block. This prevents the module from correctly recording the previous state of in after reset, leading to functional errors.

\begin{figure}[htbp]
\vspace{-1em}
\centering
\includegraphics[width=1.0\columnwidth]
{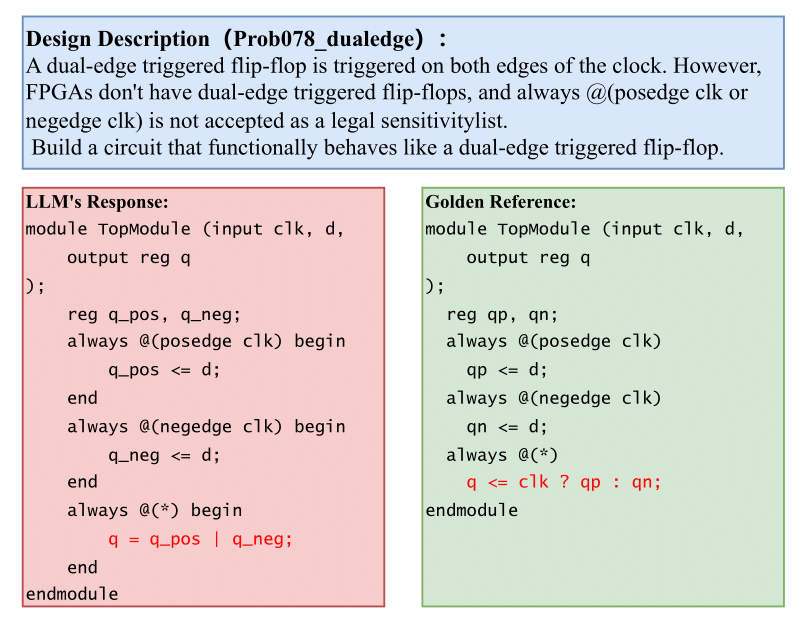}
\vspace{-1,5em}
\caption{An example of insufficient understanding of specific design}
\vspace{-1em}
\label{fig:prob078}
\end{figure}

\textbf{Insufficient Understanding of Specific Design (SP.):}
In such cases, LLMs may struggle to comprehend design specifications due to unfamiliarity with specific RTL design, leading to coding errors. This typically manifests as inadequate understanding of specialized designs such as dual-edge triggered circuits, LFSRs (Linear Feedback Shift Registers), adder carry chains, and other domain-specific implementations. As shown in Fig \ref{fig:prob078}, due to limited exposure to specialized designs like dual-edge during training, the LLM simply combines q\_pos and q\_neg through an OR operation in its final output logic, failing to fully account for the clock signal's level state to determine the correct output, thereby causing functional errors.

\begin{figure}[htbp]
\vspace{-1em}
\centering
\includegraphics[width=1.0\columnwidth]
{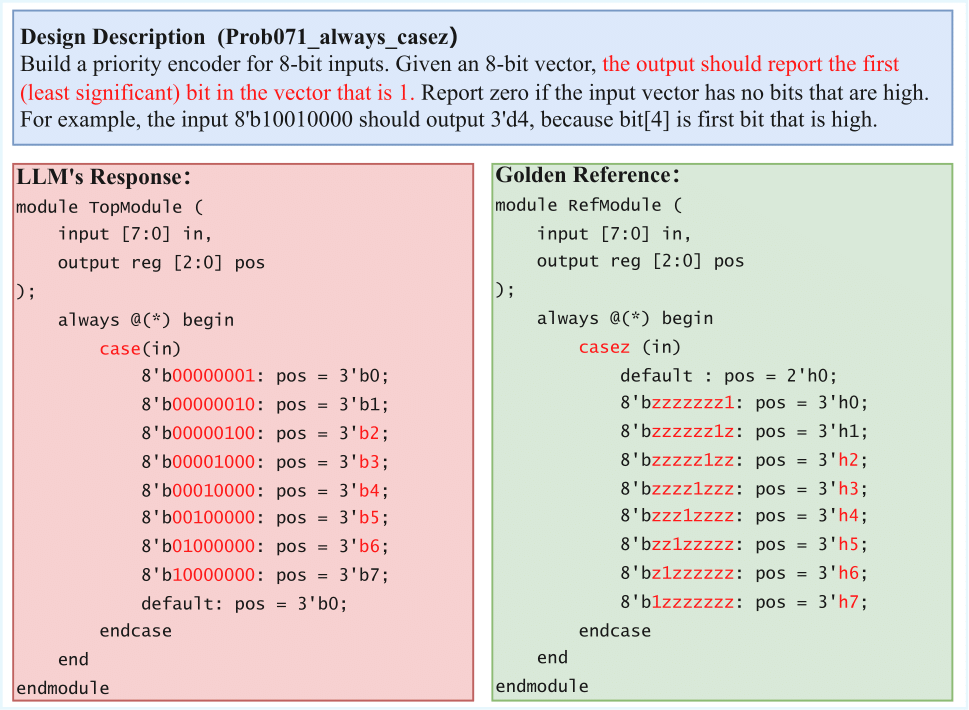}
\vspace{-1em}
\caption{An example of limitations in numerical and vector processing capabilities}
\vspace{-1em}
\label{fig:nvp}
\end{figure}

\textbf{Limitations in Numerical and Vector Processing Capabilities (NVP.):}
Additionally, due to potential limitations in its understanding of numerical and vector processing, the LLM struggles to align high-level natural language descriptions with low-level RTL code, leading to errors. Typical examples of such errors include incorrect sign-bit handling logic, misconfigured case-condition vectors, and misunderstandings of least-significant-bit (LSB) concepts—all of which stem from deficiencies in numerical and vector comprehension. 

{As shown in Fig \ref{fig:nvp}, the design description requires outputting the position of the lowest '1' bit in the input vector (e.g., for input 8'b10010000, output 4). The correct design uses casez with wildcard matching to locate this bit. However, due to a lack of knowledge in processing vectors and numbers, the LLM incorrectly uses exact case matching. This fails for inputs with multiple '1's (e.g., 8'b10010000 is misrouted to the default branch, outputting 0), resulting in a functional error.}

\begin{figure}[htbp]
\vspace{-0.8em}
\centering
\includegraphics[width=1.0\columnwidth]
{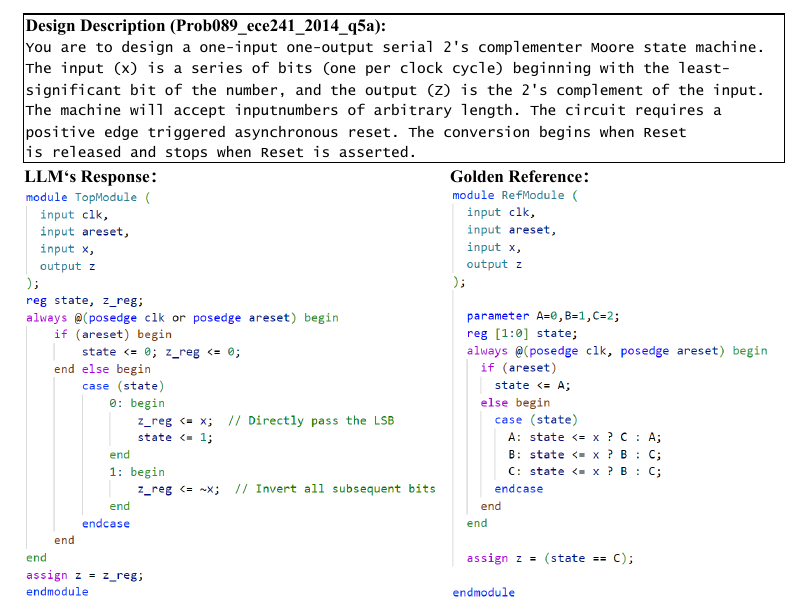}
\vspace{-1em}
\caption{An example of deficiencies in state machine design for complex scenarios}
\vspace{-1em}
\label{fig:fsm}
\end{figure}

\textbf{Deficiencies in State Machine Design for Complex Scenarios (SMDC.): }
Moreover, when LLMs process high-level functional descriptions involving state machine design—such as sequence detection problems—their limited understanding of state machine programming often results in errors in state encoding, transition conditions, and output logic. A representative example from VerilogEval is Prob089\_ece241\_2014\_q5a, as illustrated in Fig.~\ref{fig:fsm}. The specification requires a Moore machine that outputs the binary complement of the serial input. However, the LLM fails to construct an appropriate finite state machine based on this description, resulting in functional errors in the generated RTL code.

\begin{figure}[htbp]
\vspace{-0.8em}
\centering
\includegraphics[width=0.9\columnwidth]
{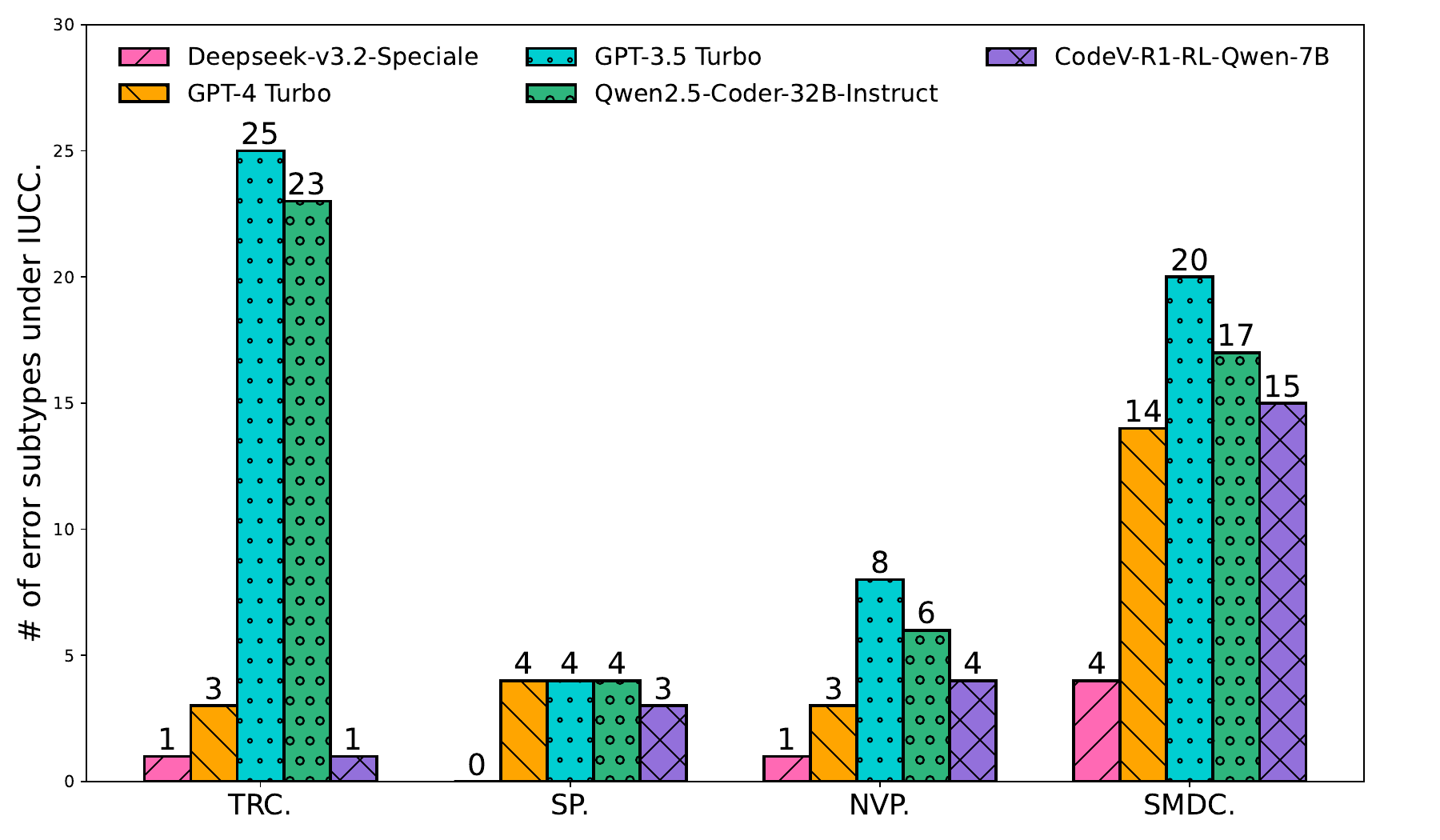}
\vspace{-1.0em}
\caption{{Distribution of error subtypes under IUCC.}}
\label{fig:iucc}
\vspace{-1em}
\end{figure}

{We have statistically analyzed the distribution of various error subtypes under the scenario of "Insufficient Understanding of Circuit Concept", as shown in Figure 6. It can be observed that, with the increase in model scale and capabilities enhanced by reasoning, \textit{DeepSeek-V3.2-Speciale} makes significantly fewer knowledge-related errors than the other four models. Meanwhile, \textit{Qwen2.5-Coder-32B-Instruct} and \textit{GPT-3.5 Turbo} exhibit notably more errors related to timing and vector/numerical processing compared to the other models. Increasing the model scale and incorporating domain-specific training data can effectively reduce these types of errors. However, all models show a relatively high number of errors in state machine design within complex circuits, indicating that this may be a key bottleneck limiting LLMs' ability to correctly generate RTL designs.}

\begin{figure}[htbp]
\vspace{-1.2em}
\centering
\includegraphics[width=1\columnwidth]
{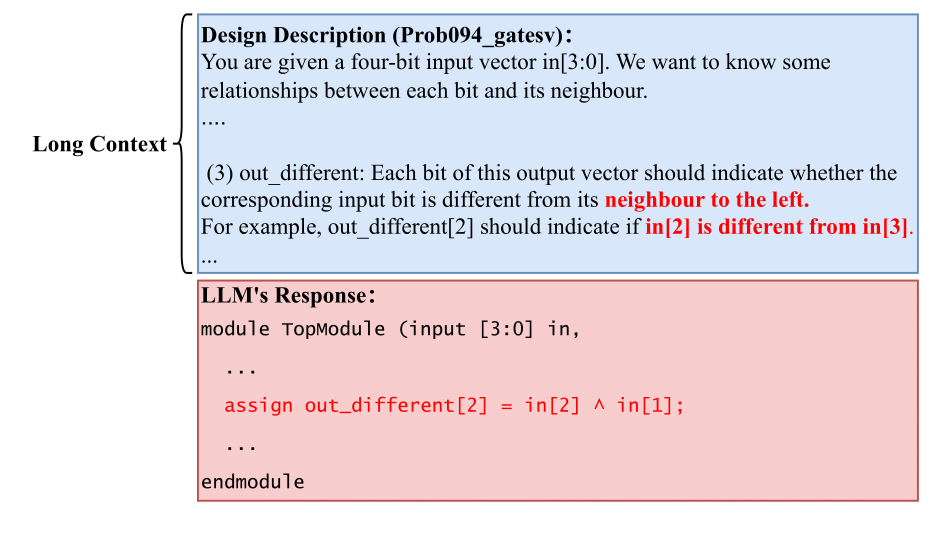}
\vspace{-1.5em}
\caption{An example of code generation errors induced by long descriptions}
\vspace{-1em}
\label{fig:prob094}
\end{figure}

\subsubsection{Missing Details of Long Descriptions (MDLD)}
Although recent LLMs claim to support long-context understanding, we observe that they often focus primarily on the main instructions in the design descriptions, while overlooking finer details. While such omissions may not significantly impact general question-answering tasks, they pose substantial challenges in LLM-based code generation, where missing even minor details can result in functional errors. As illustrated in Fig. \ref{fig:prob094}, {in the context of long input sequences, the design description requires that the \textit{i}-th bit of \texttt{out\_different} outputs the XOR value between the \textit{i}-th bit of the \texttt{in} vector and its left neighbor. However, the LLM's implementation overlooks this detail and incorrectly computes the XOR with the right neighbor instead, resulting in a functional error.}

\subsubsection{Error Distribution Statistics Under Misinteroretation of Design Specifications} 
\begin{figure}[htbp]
\vspace{-1.2em}
\centering
\includegraphics[width=0.9\columnwidth]
{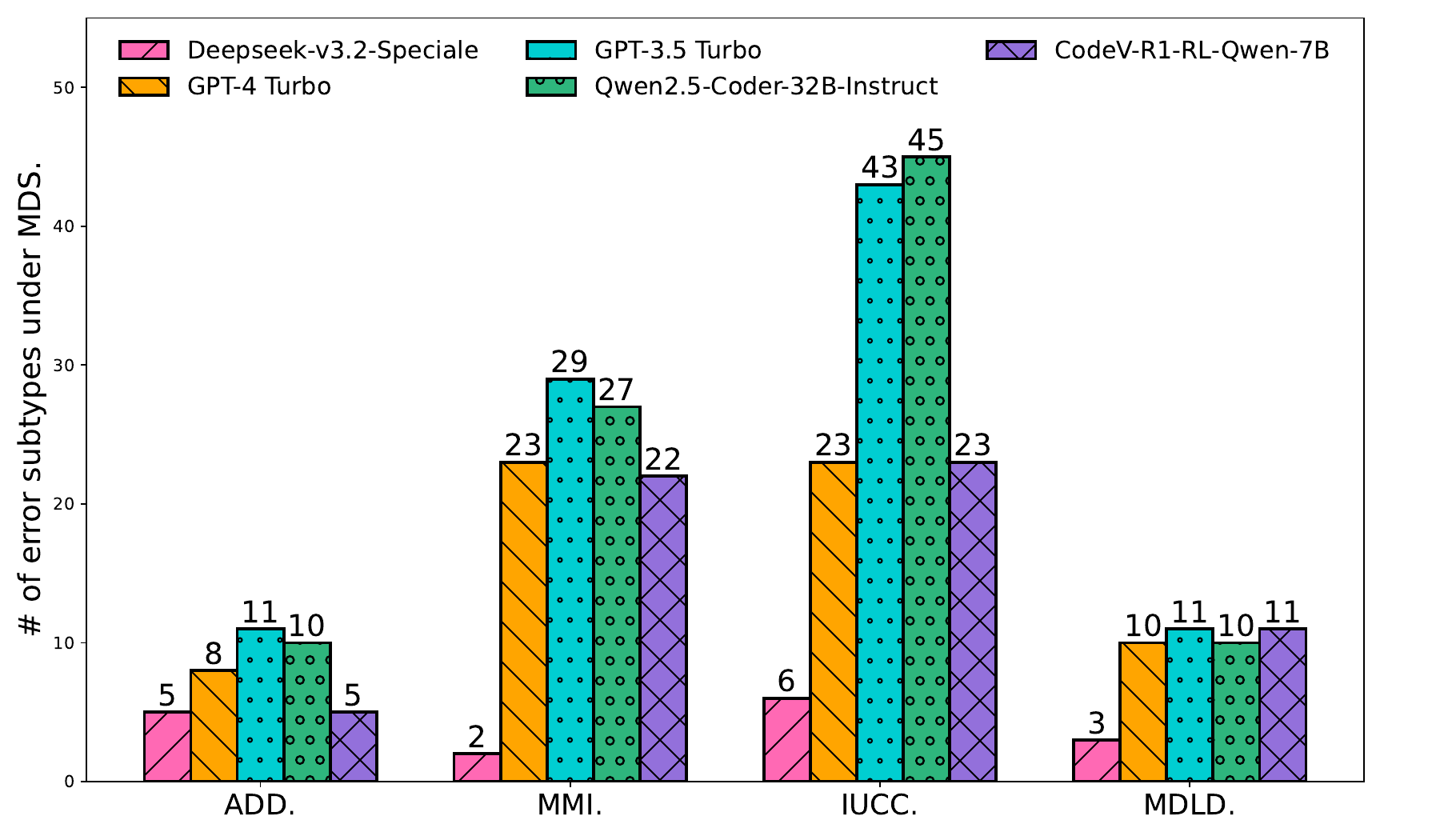}
\vspace{-1.0em}
\caption{{Errorous design distribution of error types under MDS.}}
\vspace{-1em}
\label{fig:all}
\end{figure}

We statistically analyzed the distribution of the aforementioned error types across different LLMs, as illustrated in Fig \ref{fig:all}.

Our analysis reveals that for all evaluated LLMs, the primary bottleneck in generating correct RTL code stems from insufficient circuit knowledge leading to design description misinterpretations. This is particularly pronounced in \textit{Qwen2.5-Coder-32B-Instruct} and \textit{GPT-3.5 Turbo}, where over 40 erroneous designs originated from this limitation. {GPT-4 Turbo and CodeV-R1-RL-Qwen-7B also have more than 20 designs affected by such flaws, while Deepseek-v3.2 Speciale, due to its string programming capabilities, has only 6 related errors.}


A substantial portion of errors also arises from multimodal information processing failures. The comparable performance across models in this category indicates that neither Qwen, {CodeV-R1} nor GPT series models have received adequate cross-modal alignment training specific to hardware design domains. {Thanks to its powerful comprehension and reasoning capabilities, Deepseek-v3.2-Speciale far surpasses other models in addressing such issues.}

Additionally, a smaller but consistent error category involves ambiguous descriptions and limited long-context understanding. Performance in this aspect remains similar across model scales, {from basic general-purpose models to powerful reasoning models}. Enhancing LLMs' robustness in generating RTL from vague specifications and improving their long-context comprehension in hardware design contexts remain critical research challenges.
\vspace{-1em}
\subsection{Combination of Multiple Errors}
When generating larger-scale designs, LLM-produced RTL code may contain multiple coexisting errors, rather than being limited to a single error type.
\begin{figure}[htbp]
\vspace{-1em}
\centering
\includegraphics[width=1\columnwidth]
{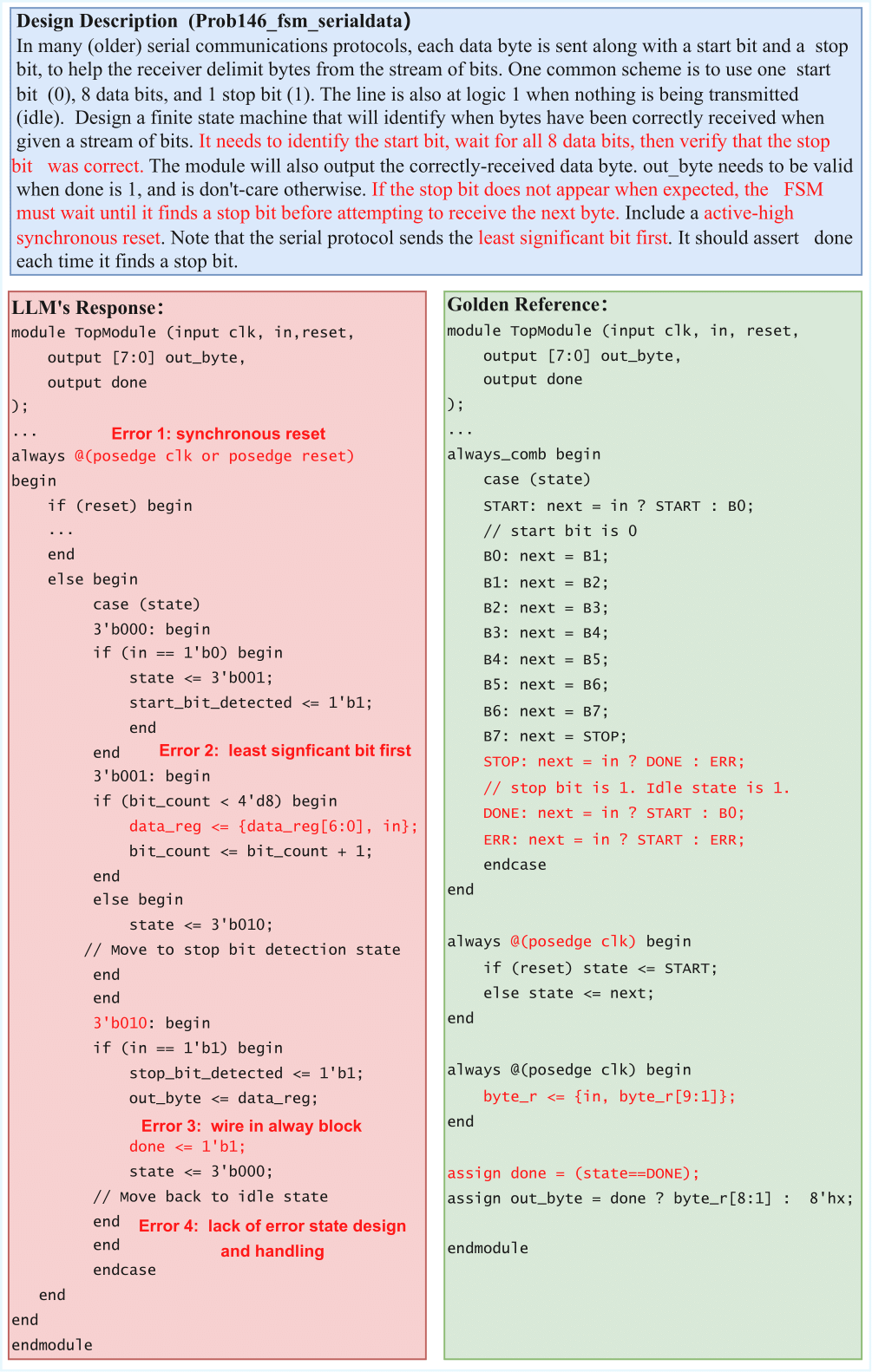}
\vspace{-1.5em}
\caption{An example of combination of multiple errors}
\vspace{-1.5em}
\label{fig:mix}
\end{figure}

\begin{itemize}
    \item \textbf{{Error 1: Inadequate Understanding of Timing Related Concepts:}}
    {The user requested the implementation of a synchronous triggering circuit, but the LLM generated an asynchronous triggering circuit.}

    \item \textbf{{Error 2: Limitations in Numerical and Vector Processing Capabilities:}}
    {The user specified Least Significant Bit first data input. To record data correctly, the data register should first store the incoming bit at the highest bit position and then shift right. However, the LLM implemented the shift in the opposite direction.}

    \item \textbf{{Error 3: Illegal Assignment to Wire in Always Block:}}
    {The LLM incorrectly assigned a value to a wire-type variable within an always block, which violates RTL programming rules. This failure is attributed to the LLM's deficient grasp of coding conventions.}

    \item \textbf{{Error 4: Deficiencies in State Machine Design for Complex Scenarios:}}
    {Upon reaching the data transmission completion state, LLM's designed state machine lacks the state transition logic for error handling.}
\end{itemize}

While some unclassified errors are categorized as miscellaneous, this work focuses on the identified ones. When LLMs understand design specifications, generated code typically aligns with golden implementations, with most errors arising from RTL programming knowledge gaps. For misinterpreted designs, causes vary—from limited circuit design understanding to ambiguous specifications or complex data formats. Certain errors, like missing long-context details, stem from inherent LLM limitations. Addressing these diverse errors requires targeted strategies to enhance LLMs' domain reasoning, disambiguation, multimodal understanding, and long-context retention.

\section{LLM-based Error Correction}

Based on the above error analysis, we propose a set of correction mechanisms tailored to address different types of errors. For issues stemming from a lack of specialized RTL programming knowledge or circuit design concepts, we generally employ Retrieval-Augmented Generation (RAG) to supplement missing knowledge using representative examples, which are then provided as context to the LLMs. To handle ambiguities in design specifications, we introduce a rule-based checking mechanism that clarifies and completes the descriptions. For some of the knowledge-related errors such as text misunderstanding that non-blocking assignments apply only to registers in sequential logic design, it is more like a design rule and is widely applied in the RTL design. In this case, we encode such domain knowledge as explicit rules and include them in the system context during code generation. To overcome challenges posed by multimodal data, we integrate external tools that assist in interpreting circuit design elements, enabling LLMs to utilize these tools for more effective RTL code generation. For errors caused by miscellaneous issues or the limited reasoning capabilities of LLMs such as omitting details in long-context scenarios, we adopt a classic iterative debugging loop that includes simulation, error localization, and correction. Finally, we integrate these error correction mechanisms into a baseline LLM-based Verilog code generation workflow to improve the overall quality of RTL code generation. The details of these mechanisms are presented in the remainder of this section.

\subsection{Error Correction Mechanisms}

\begin{figure}[htbp]
\vspace{-1em}
\centering
\includegraphics[width=0.9\columnwidth]
{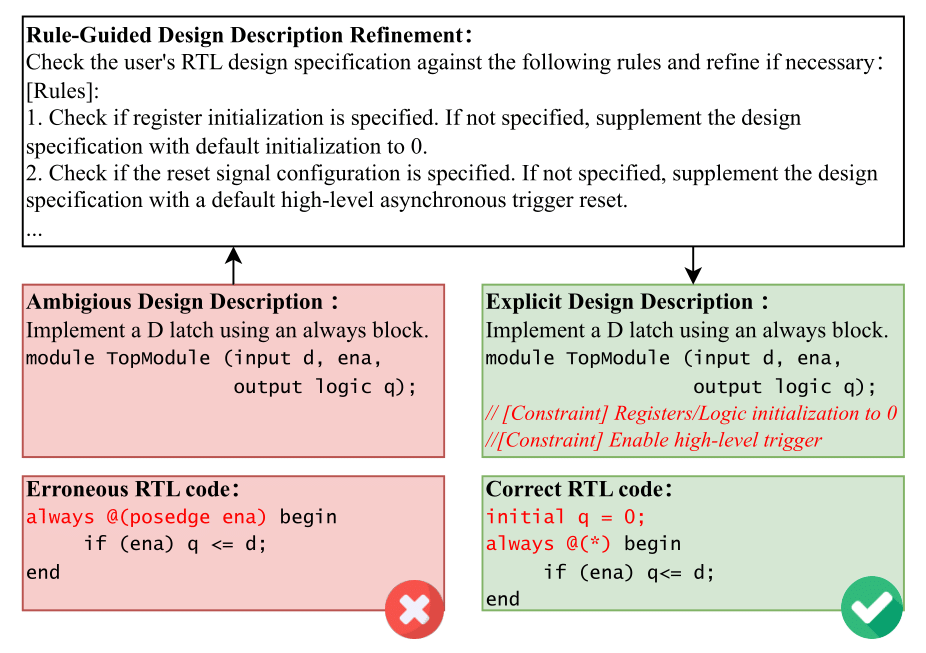}
\vspace{-1em}
\caption{An example of rule-based description refinement}
\vspace{-1.0em}
\label{fig:simple}
\end{figure}

\textbf{Rule-based Description Refinement:}
To resolve ambiguities in user‐provided specifications, we introduce a rule‑guided refinement mechanism that leverages LLMs to enforce key circuit design rules such as register initialization, reset/enable signal integrity, and internal logical consistency. The LLM first parses the initial specification, then checks it against these rules. Whenever it detects ambiguities or contradictions, the model automatically augments and clarifies the description using contextual constraints, thereby eliminating gaps in the design intent. Fig. \ref{fig:simple} illustrates a typical case: by explicitly refining the register initialization and enable signal requirements, our mechanism ensures that the subsequent Verilog code generation is both accurate and robust. Similarly, to mitigate relatively common errors arising from LLMs' unfamiliarity with specialized RTL programming, based on the analysis in Section III, we designed and integrated 10 programming rules into the system prompt, such as "Do not assign values to wire-type variables within always blocks".

\begin{figure}[htbp]
\vspace{-1em}
\centering
\includegraphics[width=0.9\columnwidth]
{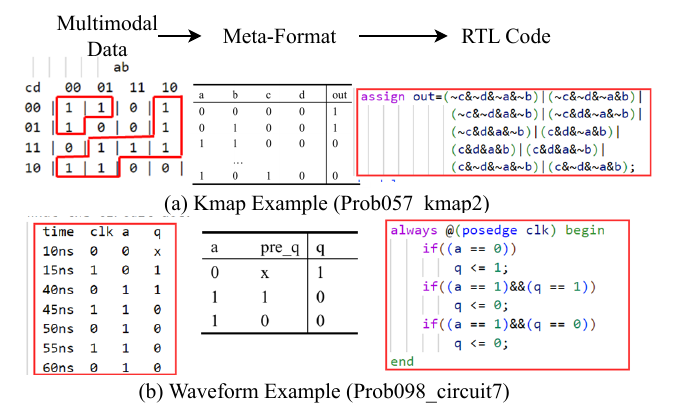}
\vspace{-0.5em}
\caption{Examples of multimodal data conversion}
\vspace{-1em}
\label{fig:multimodal_example}
\end{figure}

\textbf{Multimodal Data Conversion:}
To mitigate generation errors arising from multimodal design inputs—such as Karnaugh maps, waveform diagrams, and truth tables, we have developed tools that convert these heterogeneous formats into a single, unified truth‑table meta‑representation as shown in Fig. \ref{fig:multimodal_example}. 

{Specifically, we find that although elements such as Karnaugh maps and waveforms are difficult for LLM to understand, their inputs adhere to specific norms and standards. For instance, Karnaugh maps inputs include variable definitions in rows/columns and corresponding 0/1 arrangements, while waveforms follow a tabular format where the first column of each row contains clock information. Therefore, we first extract the multimodal information from descriptions using regular expression matching. Subsequently, we parse the extracted information based on the applicable input norms, determine the input conditions corresponding to each output item, and map them to the meta-representation of a truth table. LLMs can then translate this meta-representation into the corresponding RTL design.}



Importantly, when processing waveform inputs for sequential circuits, our conversion tool captures the clock triggering semantics to preserve correct timing behavior. Likewise, for state transition diagrams, the intermediate representation explicitly encodes both current and next states in addition to the input and output signals.

\begin{figure}[htbp]
\vspace{-1em}
\centering
\includegraphics[width=1\columnwidth]
{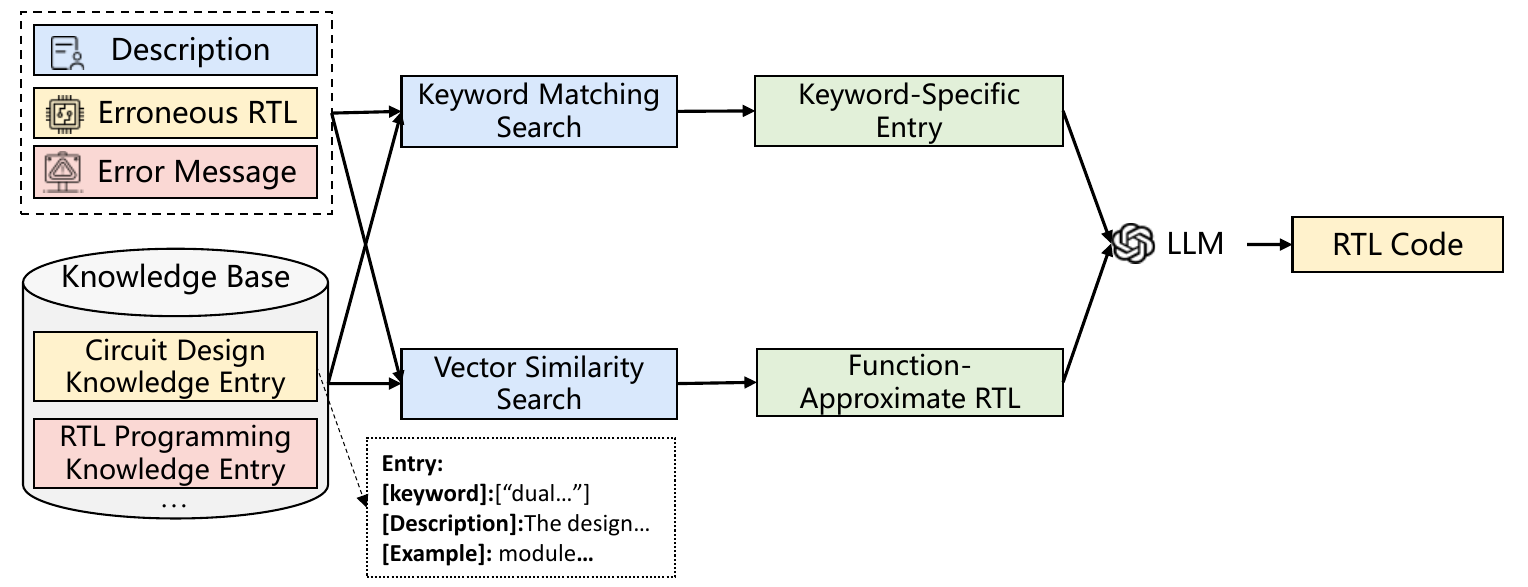}
\vspace{-1.5em}
\caption{The workflow of RAG-based knowledge error mitigation}
\vspace{-1em}
\label{fig:RAG_workflow}
\end{figure}

\textbf{RAG-based Knowledge Error Mitigation:} We leverage RAG to supplement missing knowledge in both RTL programming and circuit design, and enhance the RTL code generation quality. To this end, we construct a domain‑specific knowledge base whose entries each consist of a keyword, a concise description, and an illustrative example. 

\begin{figure}[hbtp]
\vspace{-1.0em}
\centering
\includegraphics[width=0.9\columnwidth]
{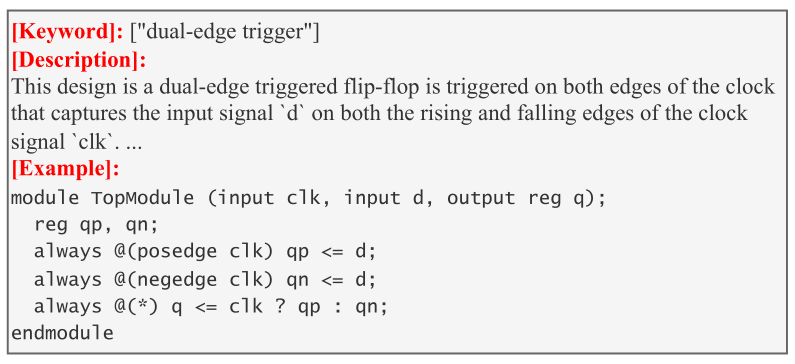}
\vspace{-0.5em}
\caption{An example of circuit design knowledge entry}
\vspace{-1em}
\label{fig:circuit_design_entry}
\end{figure}

{The knowledge base comprises two primary components. \textbf{Circuit design knowledge entries} are constructed through a three-stage pipeline: high-quality code is first collected from GitHub using keywords such as ``RTL tutorial'' and ``arithmetic circuit''; the collected repositories are then filtered by excluding those with fewer than 100 stars and heuristically removing non-functional text like licenses to reduce noise; finally, the code is annotated via few-shot learning with GPT-4 Turbo to generate natural language functional descriptions, from which relevant keywords are extracted. We have collected approximately 500 entries covering various types of circuits such as arithmetic and memory modules.
Separately, \textbf{RTL programming knowledge entries} are manually curated to address common syntax errors, with each entry providing error message, repair suggestions and concrete correction examples. We have manually designed 15 such entries. An illustrative entry is shown in Fig \ref{fig:circuit_design_entry}.}

Our retrieval engine supports both semantic and keyword‑based queries. In semantic search, user design descriptions are embedded and matched against the knowledge‑base vectors to retrieve the most relevant entries. For clearly defined RTL terms, keywords extracted from design specifications or compiler error messages are employed to retrieve corresponding knowledge entry via regular‑expression matching. Ultimately, the retrieved information is deduplicated and integrated into the LLM's context to facilitate RTL generation, thereby mitigating programming errors caused by insufficient knowledge.

\begin{figure}[htbp]
\vspace{-1em}
\centering
\includegraphics[width=0.9\columnwidth]
{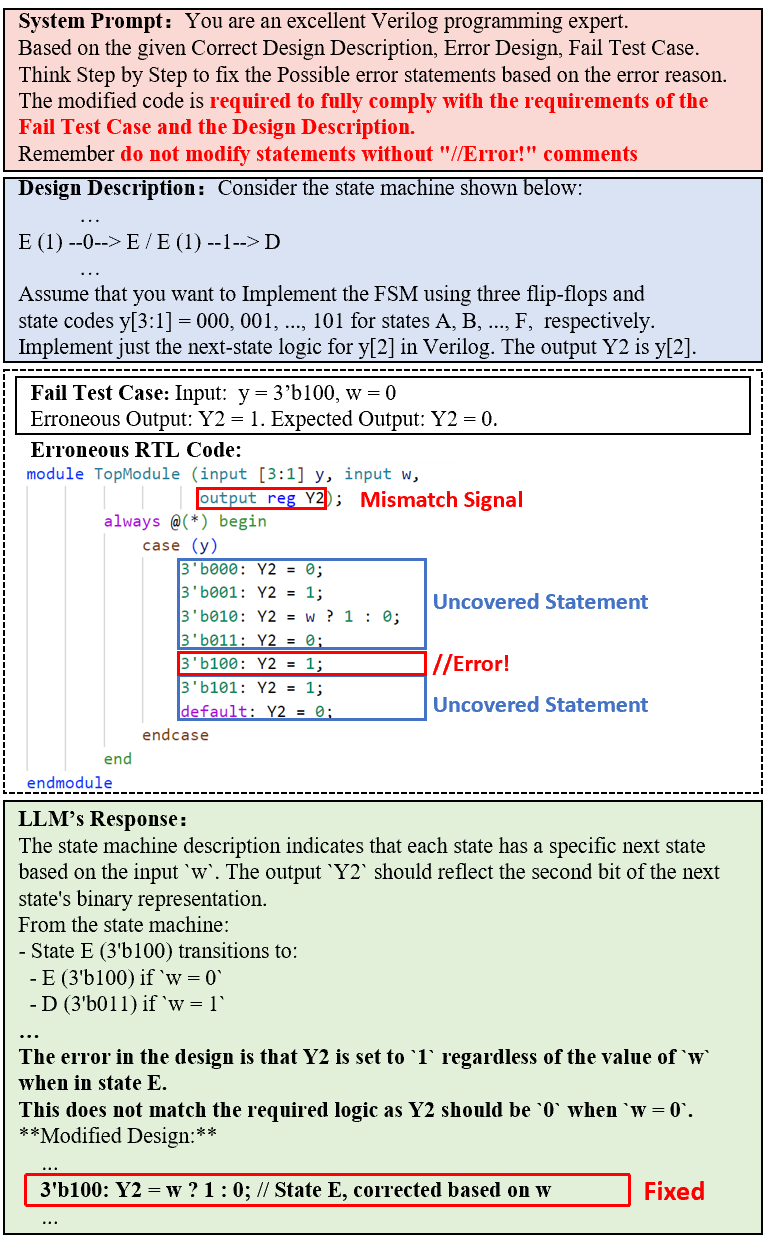}
\vspace{-0.5em}
\caption{An example of two-stage debugging mechanism }
\vspace{-1em}
\label{fig:debug_loop}
\end{figure}

\textbf{Two-Stage Debugging: Localization and Correction:}
To address the remaining errors such as missing details of long context and miscellaneous errors, we employ a simulation‑based debugging loop. First, the generated Verilog design is executed under a testbench to identify the earliest failing test case. Next, an LLM analyzes the simulation trace to pinpoint the Verilog statements most likely responsible for the failure. The failing test case, candidate error locations, and relevant design intent are then presented to the LLM, which, using chain‑of‑thought reasoning, proposes corrective modifications. This cycle of simulation, fault localization, and automated repair iterates until all test cases pass, yielding a robust final design. A representative debugging workflow is illustrated in Fig. \ref{fig:debug_loop}. In addition, specific debugging strategies can also be integrated on top of the basic debugging loop. For the missing details of long context, we may split the design descriptions into smaller semantic chunks and conduct debug of different chunks progressively.

\subsection{Application of the Error Correction Mechanisms}

\begin{figure*}[!t]
\vspace{-1em}
\centering
\includegraphics[width=0.9\textwidth]
{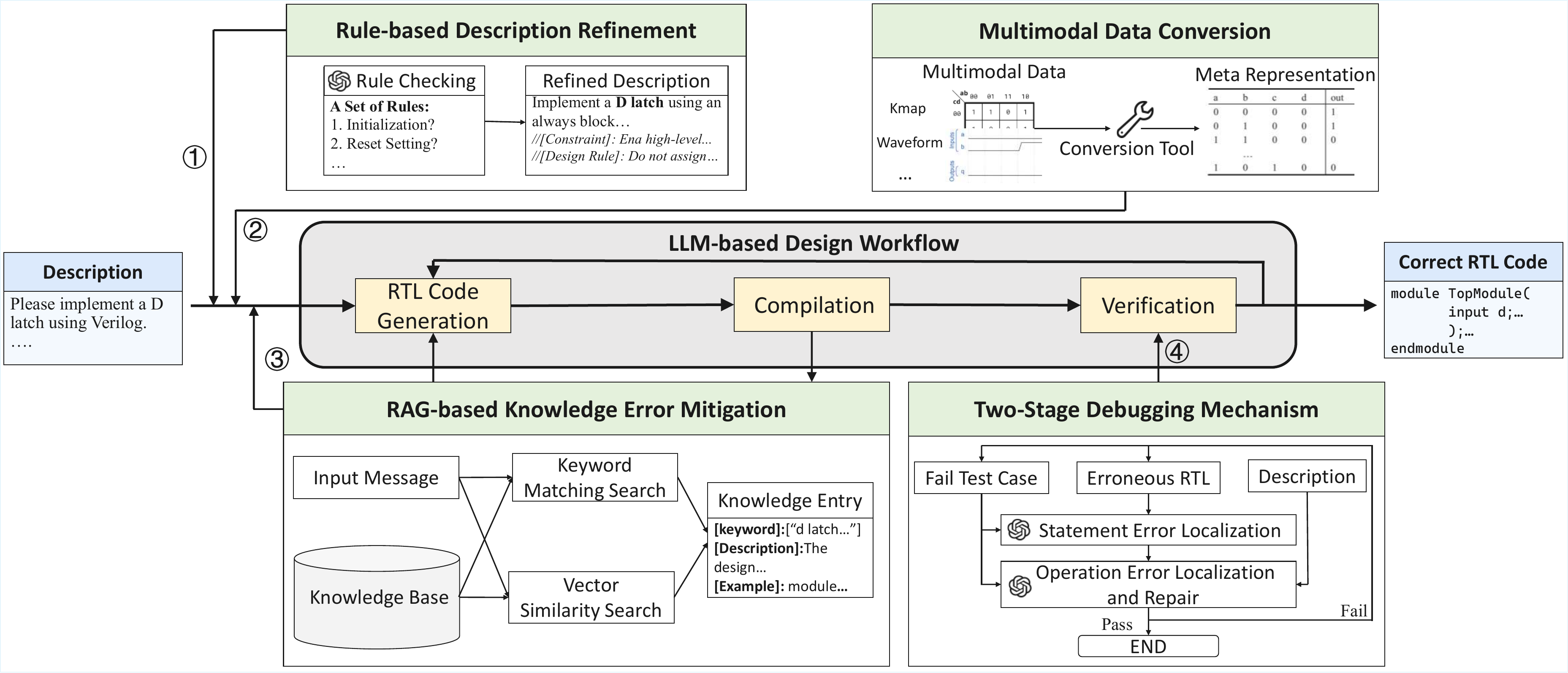}
\vspace{-0.6em}
\caption{Overview of our framework}
\vspace{-1.0em}
\label{fig:overview}
\end{figure*}

We integrate our error‑correction mechanisms into a standard LLM‑based design workflow that comprises code generation, compilation, and verification as shown in Fig. \ref{fig:overview}. 



{Given a description, the rule-based description refinement module first examines and clarifies the description while incorporating necessary RTL programming specifications to achieve initial RTL generation. If the generated RTL still contains errors, multimodal data from the description is extracted and converted via multimodal data conversion. A dedicated converter replaces such data with a unified truth-table meta-representation, simplifying subsequent code generation. If issues persist or no multimodal input is detected, our RAG-based mechanism retrieves relevant circuit designs and RTL programming knowledge from a domain-specific knowledge base to enrich the user’s design descriptions, reducing design faults due to knowledge gaps.
Finally, residual errors such as missing long-context details are exposed through simulation. Our debugger then iteratively localizes faults and invokes LLM chain-of-thought reasoning to propose and apply fixes until all test cases pass. This integrated framework ensures robust and accurate RTL generation without manual intervention.}


\section{Experiments}

\subsection{Experiment Setups}
\textbf{LLM Setups:} We evaluate Verilog code generation and repair performance using GPT-based models. All experiments are conducted with a fixed temperature of 0.1, n = 1. For the multi-round iterative RTL code repair method, we limit the maximum number of iterations to 10.

\textbf{Benchmark:} The evaluation is based on two widely used RTL repair benchmarks: \textit{VerilogEval-v1-Human} and \textit{VerilogEval-v2}, both of which contain 156 representative real-world designs spanning various difficulty levels.

\textbf{Baselines:} We compare against two categories of baselines:  
1) \textbf{Base models} — single-pass RTL code generation approaches using both General LLMs (e.g., {\textit{Deepseek-v3.2-Speciale}}), Code-Specific models (e.g. {\textit{Qwen2.5-Coder-32B-Instruct}}\cite{hui2024qwen2}) and RTL-specific models (e.g., {\textit{CodeV-R1-RL-Qwen-7B}});  
2) \textbf{Agent-based systems} — multi-agent frameworks designed to enhance LLM-based RTL generation, including
\textit{{MAGE}}~\cite{mage},
\textit{LLM+RTLFixer}~\cite{rtlfixer}, \textit{PromptV}~\cite{promptv}. 

\textbf{Methods Integration:} In our experiment, we first refine all design descriptions using rule-based method to facilitate initial RTL generation or error correction. If the design still contains errors, we perform a single-round correction via multimodal data conversion if multimodal data is detected in the design description; then we apply a single-round correction using the RAG method. If errors persist, we then adopt the two-stage debugging mechanism for iterative repair.


\vspace{-1em}
\subsection{End-to-End Code Generation Comparison}

\begin{table}[htbp] 
\small
\centering
\caption{{Main results}}
\label{tab:mainresult}
\footnotesize
\begin{tabular}{
>{\centering\arraybackslash}m{1.6cm}|
>{\centering\arraybackslash}m{2.8cm}|
>{\centering\arraybackslash}m{1.5cm}|
>{\centering\arraybackslash}m{1.5cm}}
\toprule
\textbf{Method} & \textbf{LLM Model} 
& \textbf{VerilogEval-Human (Pass@1) } 
& \textbf{VerilogEval-v2 (Pass@1) } \\

\midrule

\multirow{3}{1.8cm}{\centering Generic LLM} 
& {Deepseek-v3.2-Speciale} & {89.1} & {87.8} \\
& GPT-4 Turbo & 58.3 & 60.9\\
& GPT-3.5 Turbo & 42.9 & 41.0\\
\midrule[0.25pt]

{\centering Code-Specified LLM}
& Qwen2.5-Coder-32B-Instruct & 46.8 & 46.1 \\
\midrule[0.25pt]

\multirow{4}{1.8cm}{\centering RTL-Specified LLM} 
& RTLCoder\cite{rtlcoder} & 41.6 & 36.5  \\
& {CodeV-R1-RL-Qwen-7B}\cite{zhu2025qimeng} & {66.7} & {67.8} \\ 
& {ChipSeek-R1}\cite{chen2025chipseek} & {62.2} & {N/A} \\
\midrule[0.25pt]

\multirow{2}{1.8cm}{\centering LLM + RTLFixer\cite{rtlfixer}} 
& GPT-3.5 Turbo & 46.4 & 44.9\\
& GPT-4 Turbo & 65.0 & 65.4 \\

\midrule[0.25pt]
PromptV\cite{promptv} & GPT-4 & 80.4 & N/A \\
\midrule[0.25pt]

{MAGE}\cite{mage} & {Claude 3.5 Sonnet} & {89.1} & {93.6} \\


\midrule[0.25pt]
\multirow{5}{1.8cm}{\centering Ours}
& {Deepseek-v3.2-Speciale} & {98.1 (+9.0)} & {97.4 (+9.6)} \\
& GPT-4 Turbo & 91.0 (+32.7) & 90.4 (+29.5) \\
& GPT-3.5 Turbo & 73.1 (+30.2) & 72.4 (+31.4)\\
& Qwen2.5-Coder-32B-Instruct & 76.9 (+30.1) & 78.1 (+32.0) \\
& {CodeV-R1-RL-Qwen-7B} & {89.7 (+23.0)} & {91.7 (+23.9) }\\

\bottomrule
\end{tabular}
\vspace{-2.70em}

\end{table}

Table~\ref{tab:mainresult} presents a comparison between our method and existing baseline approaches. Our solution achieves \textbf{{98.1\%}} and \textbf{{97.4\%}} accuracy in pass@1 evaluations on \textit{VerilogEval-Human} and \textit{VerilogEval-v2}, respectively. 
{As can be seen, our method delivers performance improvements ranging from approximately 9\% to 30\%, whether applied to the powerful general-purpose model DeepSeek-v3.2-Speciale or to smaller models additionally trained on Code and RTL.}
In addition, our approach outperforms various agent-based systems including \textit{MAGE},\textit{LLM+RTLFixer}, and \textit{PromptV}. Meanwhile, 
{by integrating our code generation and repair framework, CodeV-R1-RL-Qwen-7B achieves performance comparable to the current SOTA agent framework, MAGE. Meanwhile, based on DeepSeek-v3.2-Speciale, our framework demonstrates performance improvements over MAGE of 9.0\% on VerilogEval-Human and 4.1\% on VerilogEval-v2, respectively.}



\vspace{-1em}
\subsection{Error Correction Analysis}

\begin{figure}[htbp]
\vspace{-1em}
\centering
\includegraphics[width=0.9\columnwidth]
{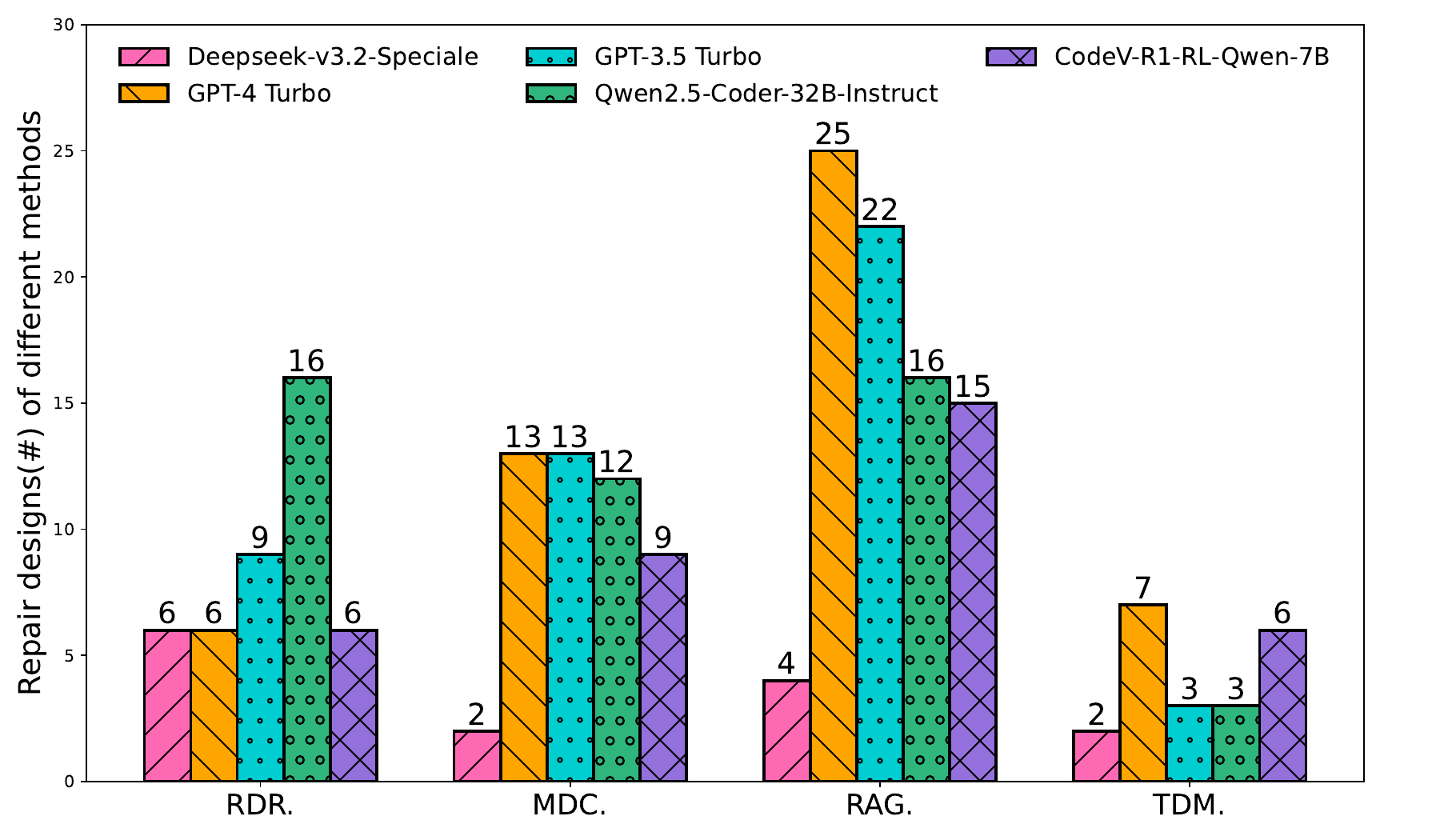}
\vspace{-1.0em}
\caption{{Number of correctly repaired designs by different methods }}
\vspace{-1em}
\label{fig:ablation_study}
\end{figure}

When utilizing foundational LLMs to generate code on the VerilogEval-Human benchmark, 
we use each model to self-correct their own erroneous designs, as shown in Table \ref{tab:model_errors}. The number of successfully repaired designs for each method is summarized in Fig~\ref{fig:ablation_study}.

\textbf{Impact of Rule-based Description Refinement (RDR.):} 
We observe that some design descriptions are vague, such as missing explanations of input signals. Additionally, we identify several issues stemming from illegal assignments. {It can be seen that all models based on this module can fix more than 6 faulty designs, and in particular,\textit{ Qwen2.5-Coder-32B-Instruct} can fix 16 designs.}


\textbf{Impact of Multimodal Data Conversion (MDC.):} 
Our experiments reveal that current LLMs encounter challenges in semantically understanding multimodal hardware design data. {It can be seen that all models based on this module—except for \textit{Deepseek-v3.2-Speciale} model—can correctly fix 9 or more designs. Moreover, integrating this module into \textit{Deepseek-v3.2-Speciale} can also fully fix all errors caused by misunderstandings of Waveform.}


\textbf{Impact of RAG-based Knowledge Error Mitigation (RAG.):} 
Experimental results demonstrate that this method significantly reduces knowledge-related errors in RTL designs generated by GPT models. {Meanwhile, this module can also reduce 15 faults caused by deficiencies in circuit knowledge mastery in \textit{Qwen2.5-Coder-32B-Instruct} and \textit{CodeV-R1}. On \textit{Deepseek-v3.2-Speciale}, this module can also fix approximately 4 faulty designs.}

\textbf{Impact of Two-Stage Debugging Mechanism (TDM.):} 
We find that the debugging module performs better with \textit{GPT-4 Turbo} and {\textit{CodeV-R1}} than with \textit{GPT-3.5 Turbo} and \textit{Qwen2.5-Coder-32B-Instruct}. This may be attributed to models like \textit{GPT-3.5 Turbo} tendency to produce larger and more error-prone code blocks that require extensive revisions. 

\begin{table*}[htbp] 
\small
\centering
\footnotesize
\caption{{Correction outcomes for different categories of errors}}
\label{tab:fix_study}

\begin{tabular}{
>{\centering\arraybackslash}m{2.8cm}|
>{\centering\arraybackslash}m{0.5cm}|
>{\centering\arraybackslash}m{0.5cm}|
>{\centering\arraybackslash}m{0.7cm}|
>{\centering\arraybackslash}m{0.5cm}|
>{\centering\arraybackslash}m{0.5cm}|
>{\centering\arraybackslash}m{0.7cm}|
>{\centering\arraybackslash}m{0.5cm}|
>{\centering\arraybackslash}m{0.5cm}|
>{\centering\arraybackslash}m{0.7cm}|
>{\centering\arraybackslash}m{0.5cm}|
>{\centering\arraybackslash}m{0.5cm}|
>{\centering\arraybackslash}m{0.7cm}|
>{\centering\arraybackslash}m{0.5cm}|
>{\centering\arraybackslash}m{0.5cm}|
>{\centering\arraybackslash}m{0.7cm}
}
\toprule
\multirow{2}{2.8cm}{\centering \textbf{Model}}  
& \multicolumn{3}{c|}{\textbf{IKSP.}}  
& \multicolumn{3}{c|}{\textbf{ADD.}}  
& \multicolumn{3}{c|}{\textbf{MMD.}}
& \multicolumn{3}{c|}{\textbf{IUCC.}}
& \multicolumn{3}{c}{\textbf{MDLD.}}
\\
\cmidrule[0.25pt]{2-16}
~ 
& \textbf{Total} & \textbf{Pass} & \textbf{Acc} 
& \textbf{Total} & \textbf{Pass} & \textbf{Acc}
& \textbf{Total} & \textbf{Pass} & \textbf{Acc} 
& \textbf{Total} & \textbf{Pass} & \textbf{Acc} 
& \textbf{Total} & \textbf{Pass} & \textbf{Acc} 
\\
\midrule
{Deepseek-v3.2-Speciale}
& 2 & 2 & 100.0
& 5 & 4 & 80.0
& 2 & 2 & 100.0
& 6 & 5 & 83.3
& 3 & 2 & 66.7
\\
\midrule[0.25pt]
GPT-4 Turbo
& 8 & 7 & 87.5
& 8 & 4 & 50.0
& 23 & 17 & 73.9 
& 23 & 18 & 78.3 
& 10 & 4 & 40.0 
\\
\midrule[0.25pt]
GPT-3.5 Turbo
& 25 & 14 & 56.0
& 11 & 3 & 27.3
& 28 & 19 & 67.9
& 43 & 18 & 41.9
& 11 & 5 & 45.5
\\
\midrule[0.25pt]
Qwen2.5-Coder-32B-Instruct
& 13 & 8 & 61.5 
& 10 & 3 & 30.0 
& 27 & 18 & 66.7
& 45 & 23 & 51.1
& 10 & 1 & 10.0
\\

\midrule[0.25pt]
{CodeV-R1-RL-Qwen-7B}  &
3 & 2 & 66.7 &
5 & 3 & 60.0 &
23 & 13 & 56.5 &
22 & 16 & 72.7 &
11 & 4 & 36.4
\\

\bottomrule
\end{tabular}
\vspace{-0.8em}
\begin{flushleft}
\footnotesize
\textsuperscript{*}Acc: Accuracy (\%), calculated as Pass/Total × 100. \\
\end{flushleft}
\vspace{-3em}

\end{table*}


Subsequently, we investigated the effectiveness of error correction across different error categories, based on 243 erroneous RTL designs generated by the three different LLMs above. 
The correction outcomes for each error category are summarized in Table~\ref{tab:fix_study}.


It can be observed that for cases of Insufficient Knowledge of Specialized RTL Programming, the repair results are notably effective, On all models, this method can fix over 60\% of such incorrectly designed cases.
When user descriptions are ambiguous, the framework can still rectify a subset of resulting errors, contributing to more robust code generation. {For Misinterpretation of Multimodal Data, our framework can fix more than 13 incorrectly designed cases on all models except \textit{Deepseek-v3.2-Speciale}. Similarly, for Insufficient Understanding of Circuit Concepts, our framework can fix over 16 incorrectly designed cases caused by this issue. Furthermore, in complex long-context scenarios, it can be observed that when our framework is based on general-purpose LLMs, its repair performance is better compared to code-specific LLMs—it can fix over 40\% of the designs with faults caused by this scenario.}


We have found that complex design concepts combining multiple knowledge components, such as the gshare branch predictor with multiple design considerations, cannot guarantee correct RTL generation based solely on RAG. Similarly, problems like Prob093\_ece241\_2014\_q3—which require implementing a given K-map logic function using a 4-to-1 multiplexer and minimal 2-to-1 multiplexers—demand deeper multimodal data understanding to meet complex design constraints, and intermediate representation remains limited in such cases. Additionally, LLMs' lack of detailed long-context understanding further limits the correctness of RTL generation. Enhancing LLMs' understanding and analytical capabilities for such scenarios represents a key challenge and future research direction.

\section{Discussion}
In this section, we will illustrate our error classification criteria through several representative examples.

\vspace{-1em}
\subsection{Classification criteria for the two main error categories}

When given natural language design descriptions, we consider an LLM to have correctly understood the requirements if it generates accurate high-level pseudocode. We conclude the LLM failed to comprehend the design requirements when its generated pseudocode exhibits behavioral inconsistencies with the specifications. Conversely, if the LLM violates RTL syntax rules when translating correct pseudocode into low-level RTL code, we attribute this to insufficient mastery of RTL programming knowledge.

\begin{figure}[htbp]
\vspace{-1em}
\centering
\includegraphics[width=1.0\columnwidth]
{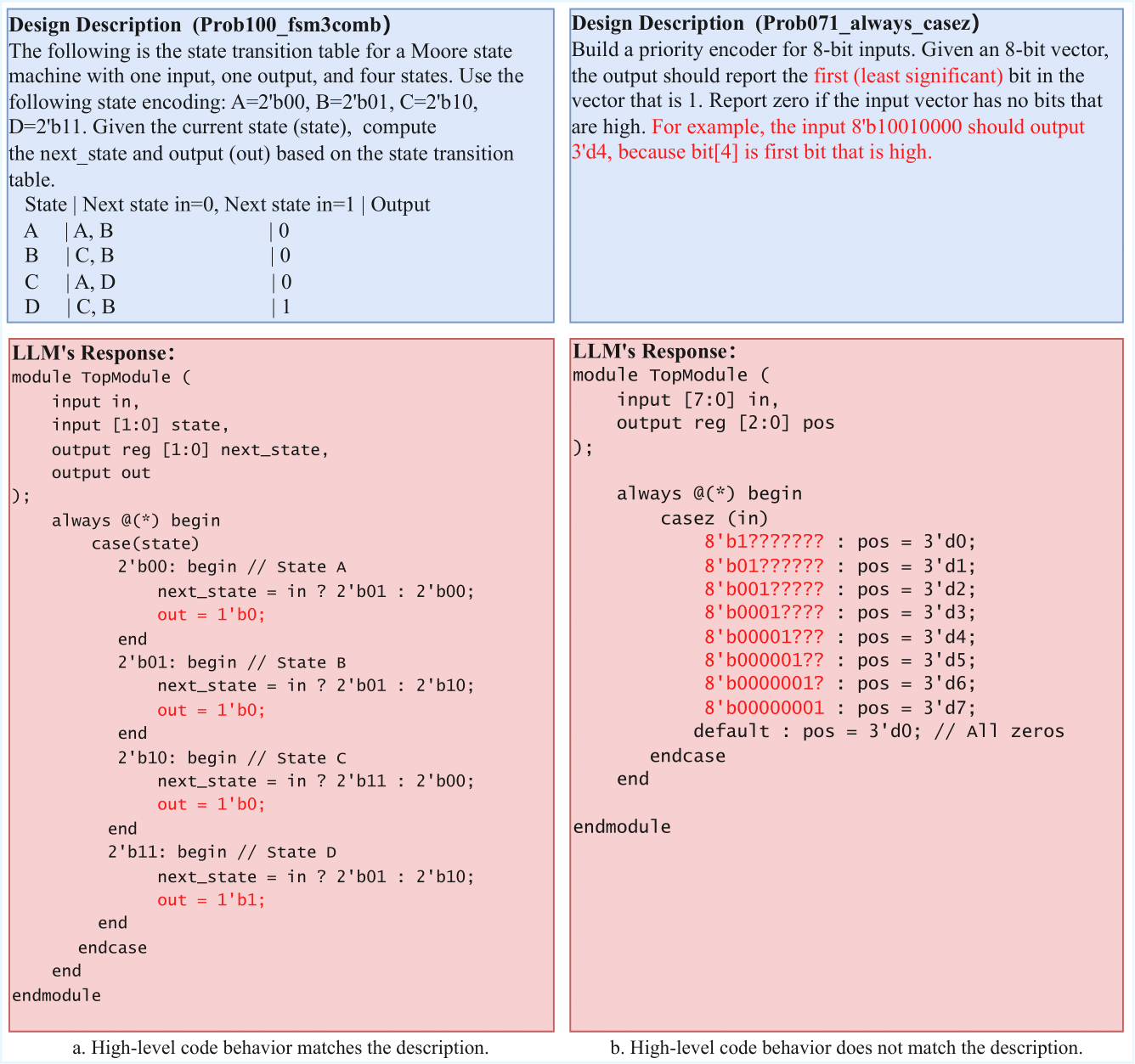}
\vspace{-1em}
\caption{Two distinct error type examples}
\label{fig:cate_1}
\vspace{-1em}
\end{figure}

As illustrated in Fig. \ref{fig:cate_1}, Sample a demonstrates correct high-level code behavior (state design, transitions, and output logic) that complies with the description, but contains RTL implementation errors due to improper wire-type variable assignment within an always block. Sample b exhibits behavioral inconsistencies with the description resulting from the LLM's incorrect understanding of the least significant bit concept - representing two distinct error categories.

Notably, some larger designs may contain both error types simultaneously (Fig. \ref{fig:mix}), 
such designs are classified into both categories.

\vspace{-1em}
\subsection{Subclassification Criteria for Misinterpretation of Design Specifications}

\begin{figure}[htbp]
\vspace{-1em}
\centering
\includegraphics[width=1.0\columnwidth]
{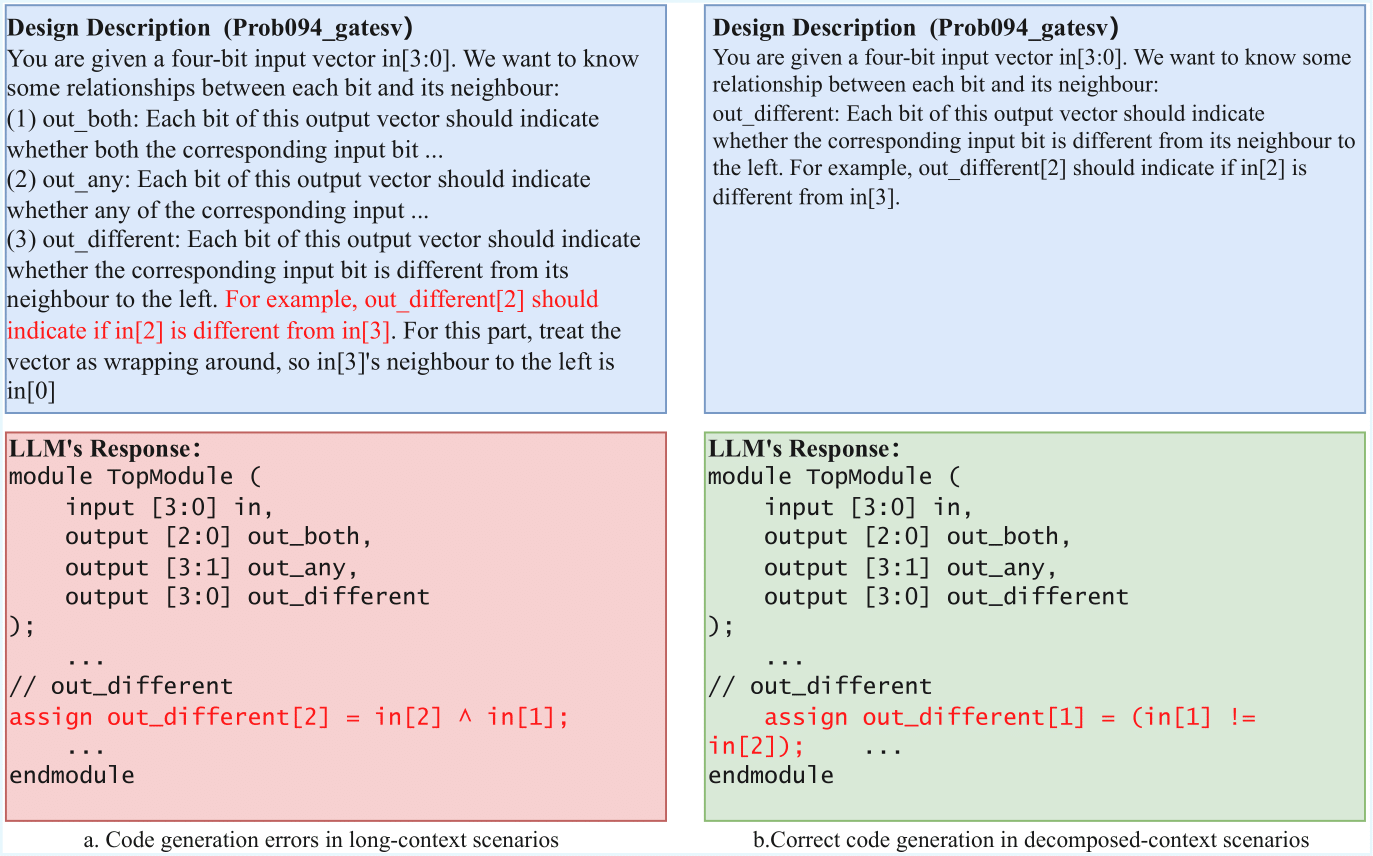}
\vspace{-1em}
\caption{Long-context constraints hinder LLMs from correctly generating RTL code examples}
\vspace{-1.5em}
\label{fig:cate_2}

\end{figure}

In long-context scenarios, directly prompting LLMs to generate RTL code from lengthy descriptions often leads to inaccuracies in certain code segments. However, manually decomposing the context—keeping only key requirements and removing unnecessary details (Fig \ref{fig:cate_2})—enabled the LLM to produce correct RTL code. This shows that redundancy in long contexts limits the accuracy of LLM-generated RTL code.

\begin{figure}[htbp]
\vspace{-1em}
\centering
\includegraphics[width=1.0\columnwidth]
{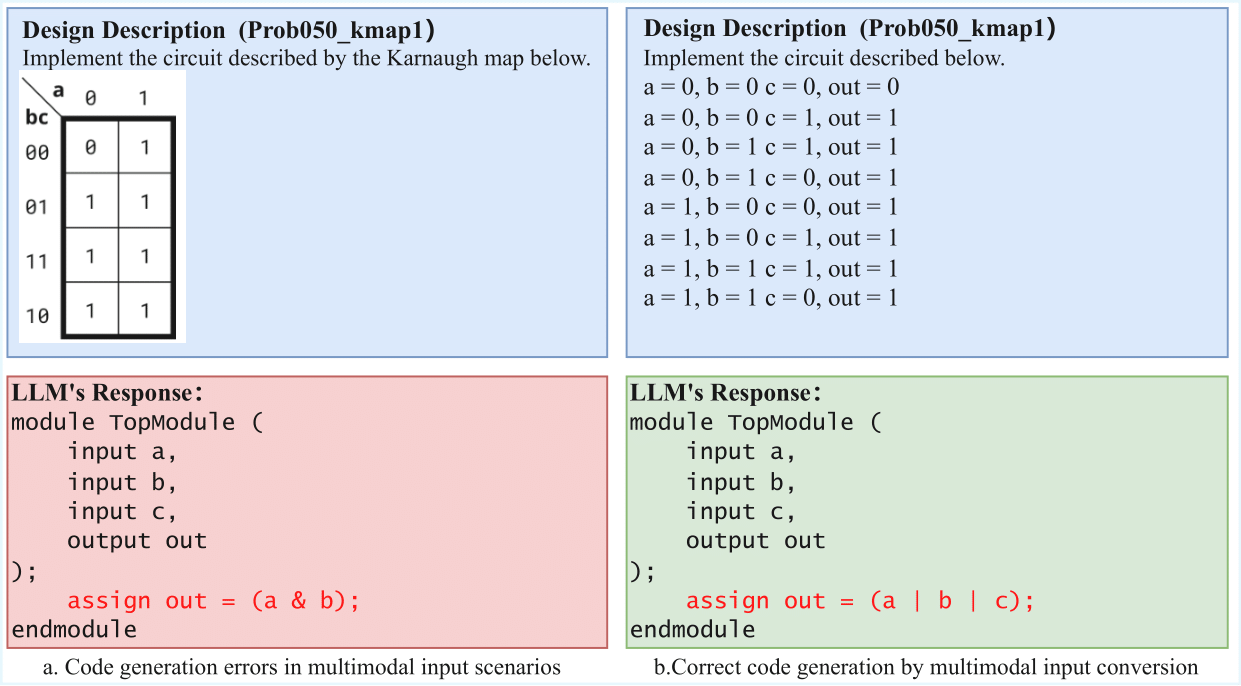}
\vspace{-1em}
\caption{Multimodal input constraints hinder LLMs from correctly generating RTL code examples}
\vspace{-1.0em}
\label{fig:cate_3}
\end{figure}


Similarly, as shown in Fig. \ref{fig:cate_3}, when the content of a KMAP table is converted into natural language, the LLM successfully generates functionally accurate RTL code, revealing a bottleneck in its ability to directly comprehend tabular or other non-textual modalities.

\begin{figure}[htbp]
\vspace{-1em}
\centering
\includegraphics[width=1.0\columnwidth]
{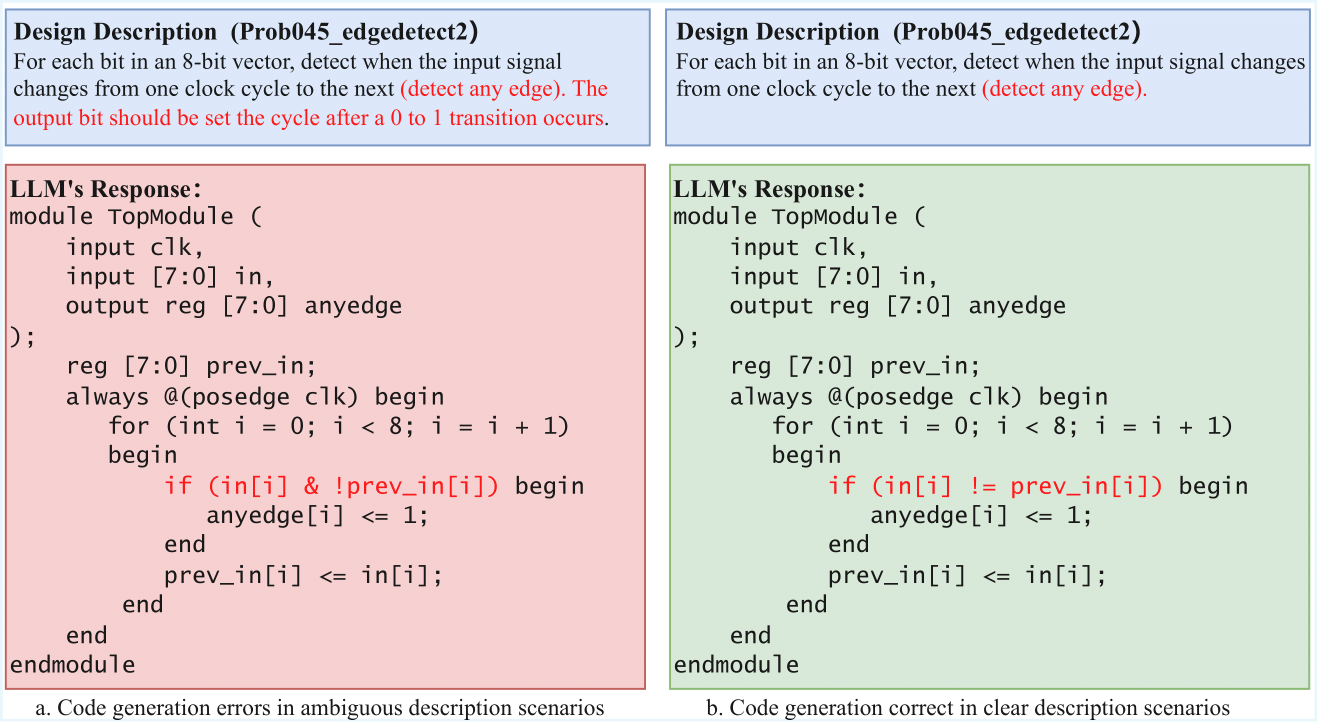}
\vspace{-0.5em}
\caption{Ambiguous description constraints hinder LLMs from correctly generating RTL code examples}
\vspace{-1em}
\label{fig:cate_4}
\end{figure}

\begin{figure}[htbp]
\vspace{-1em}
\centering
\includegraphics[width=1.0\columnwidth]
{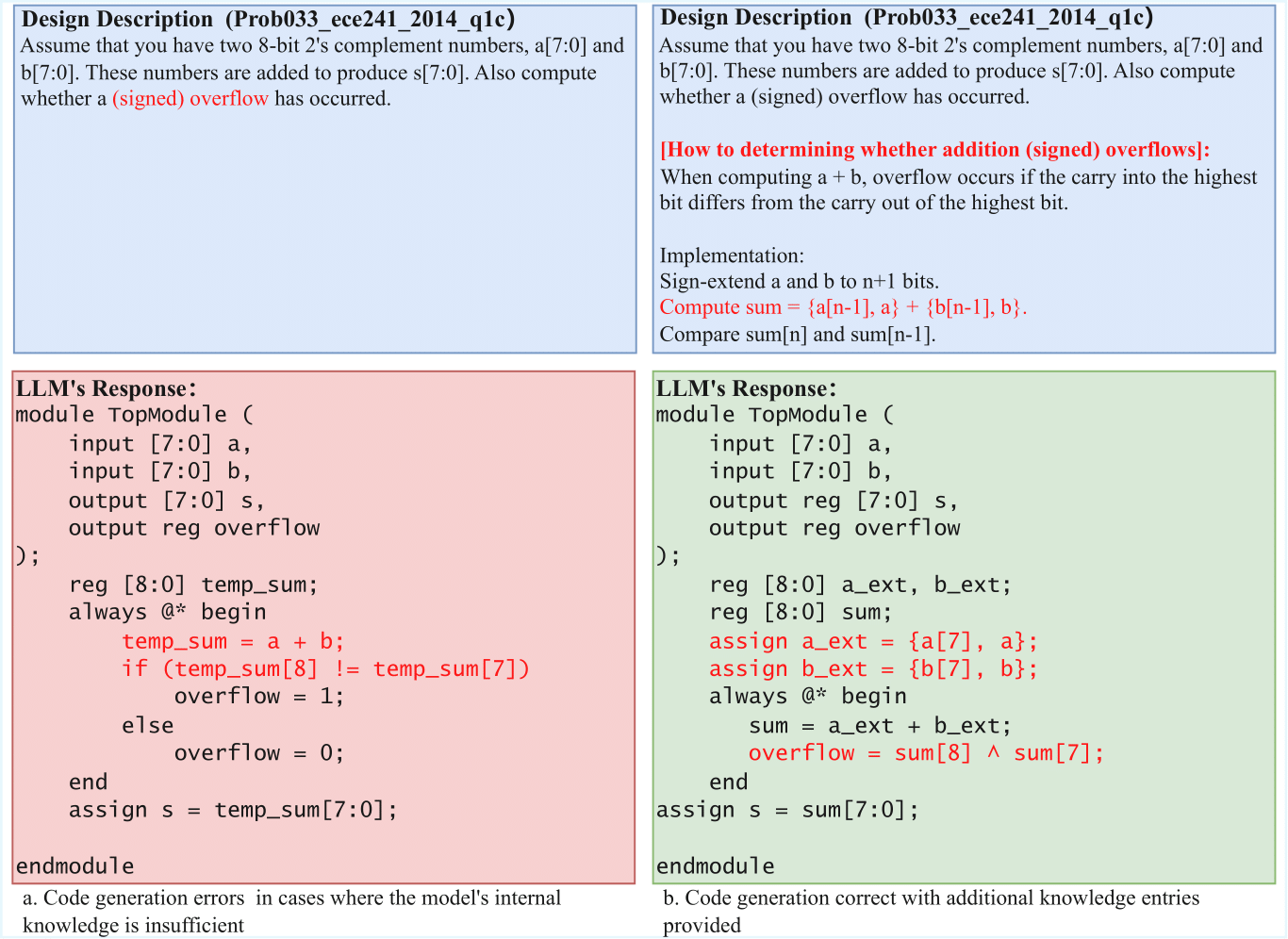}
\vspace{-1em}
\caption{Insufficient mastery of circuit knowledge constraints hinder LLMs from correctly generating RTL code examples}
\vspace{-1.5em}
\label{fig:cate_5}
\end{figure}




As shown in Fig. \ref{fig:cate_4}, ambiguous design descriptions required manual refinement by removing contradictions and supplementing key information. This clarification significantly improved RTL generation accuracy. Remaining errors primarily stemmed from the LLM's limited circuit knowledge, causing RTL-description mismatches. Cases like Fig. \ref{fig:cate_5} demonstrate that providing essential circuit knowledge in the context helps generate correct RTL, confirming this knowledge gap as a major limitation.


\section{Conclusion}
In this paper, we systematically analyzed and categorized error causes in LLM-generated RTL code, and find that most errors stem not from the reasoning capabilities of LLMs, but rather from a lack of RTL programming knowledge, insufficient understanding of circuit concepts, ambiguous design descriptions, or misinterpretation of complex multimodal inputs. To address these, we propose a set of error correction techniques including: 1) rule-based description refinement, 2) multimodal data conversion,  3) RAG-based knowledge error mitigation, and 4) simulation-guided iterative debugging. We integrate the proposed error correction mechanisms into a representative LLM-based RTL code generation framework, achieving  98.1\% accuracy on the VerilogEval Benchmark. This demonstrates that based on a series of strategies such as knowledge retrieval, without requiring additional training, can significantly mitigate errors in LLM-generated RTL code.

\bibliographystyle{unsrt}
\bibliography{ref}
\vspace{-5em}

\begin{IEEEbiography}[{\includegraphics[width=1in,height=1.25in,clip,keepaspectratio]{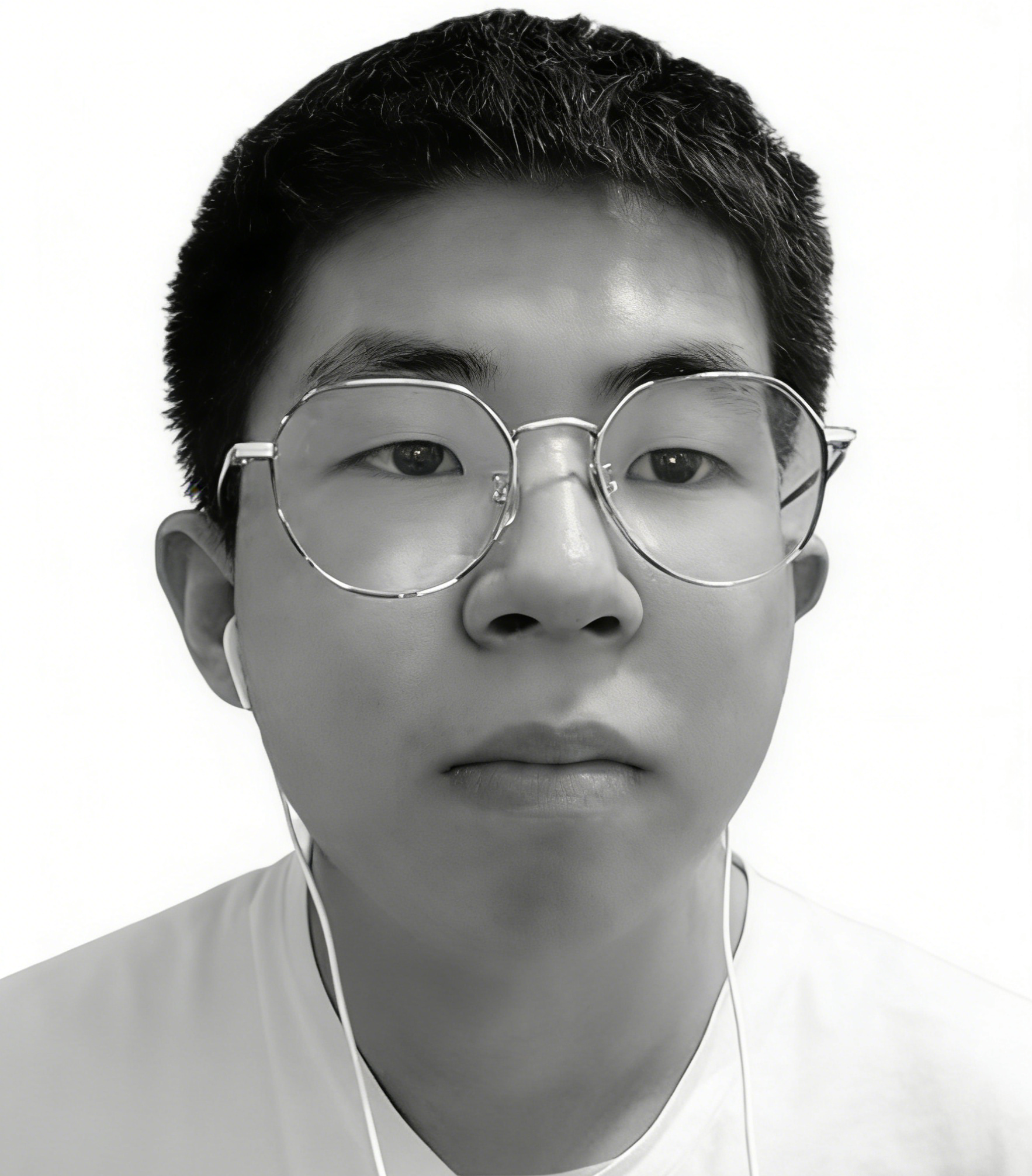}}]{Jiazheng Zhang}
received the B.Eng. degree in computer science and technology from  Harbin Institute of Technology, Weihai, China,
in 2024. He is currently pursuing the M.Eng. degree
with the Institute of Computing Technology, Chinese
Academy of Sciences, Beijing, China, and the
Department of Computer Science, University of
Chinese Academy of Sciences, Beijing, China.

His currently research falls primarily in the field
of LLM for Chip Design and LLM for RTL generation.

\end{IEEEbiography}
\vspace{-5em}
\begin{IEEEbiography}[{\includegraphics[width=1in,height=1.25in,clip,keepaspectratio]{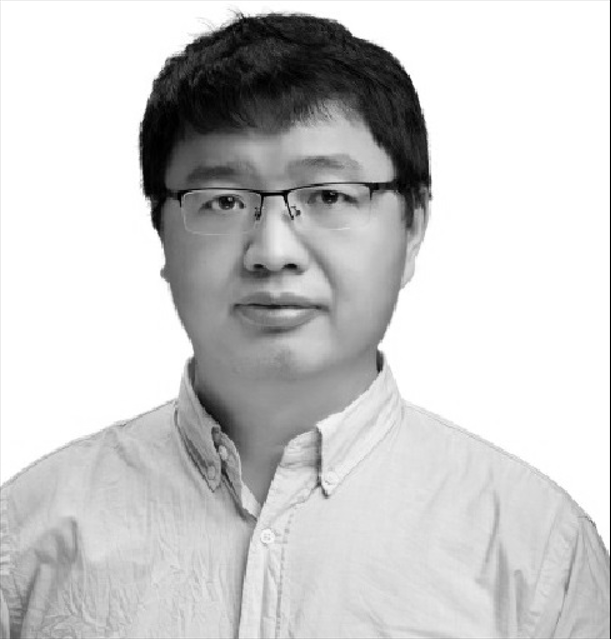}}]{Cheng Liu}
(Senior Member, IEEE) received the B.Eng. and M.Eng. degrees from the Harbin Institute of Technology, Harbin, China, in 2007 and 2009, respectively, and the Ph.D. degree from the University of Hong Kong, Hong Kong, in 2016. 

He is currently an Associate Professor with the Institute of Computing Technology, Chinese Academy of Sciences, Beijing, China. His research interests include domain specific architecture design and LLM-assisted chip design.
\end{IEEEbiography}
\vspace{-5em}
\begin{IEEEbiography}[{\includegraphics[width=1in,height=1.25in,clip,keepaspectratio]{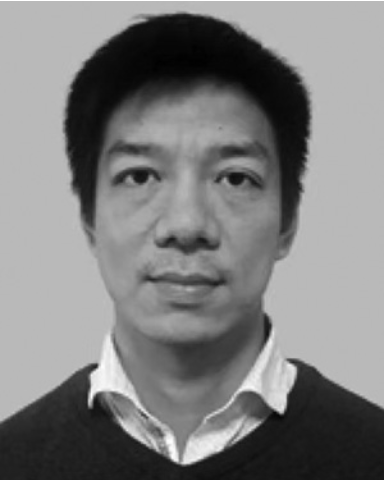}}]{Long Cheng}
(Senior Member, IEEE) received the B.E. degree from Harbin Institute of Technology, China, in 2007, the M.Sc. degree from the University of Duisburg–Essen, Germany, in 2010, and the Ph.D. degree from Maynooth University, Ireland, in 2014.

He is a full professor at the School of Control and Computer Engineering, North China Electric Power University, Beijing. Previously, he was an assistant professor at Dublin City University and a Marie Curie Fellow at University College Dublin. His research focuses on distributed systems, deep learning, and cloud computing.

\end{IEEEbiography}
\vspace{-4em}
\begin{IEEEbiography}[{\includegraphics[width=1in,height=1.25in,clip,keepaspectratio]{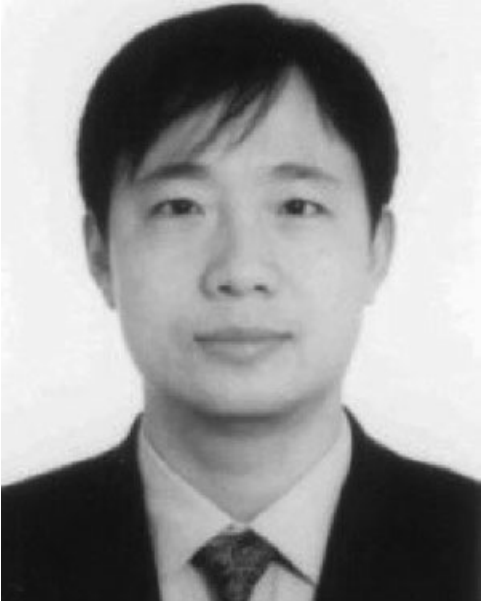}}]{Xiaowei Li}
 (Senior Member, IEEE) received the
B.Eng. and M.Eng. degrees in computer science
from the Hefei University of Technology, Hefei,
China, in 1985 and 1988, respectively, and the Ph.D.
degree in computer science from the Institute of
Computing Technology (ICT), Chinese Academy of
Sciences (CAS), Beijing, China, in 1991.

He was an Associate Professor at Peking University (1991–2000) and became a Professor at the Institute of Computing Technology, Chinese Academy of Sciences (ICT, CAS) in 2000. He has co-authored over 280 papers, holds 60 patents and 30 software copyrights, and his research focuses on VLSI testing, design verification, and dependable computing.

\end{IEEEbiography}
\vspace{-4em}
\begin{IEEEbiography}[{\includegraphics[width=1in,height=1.25in,clip,keepaspectratio]{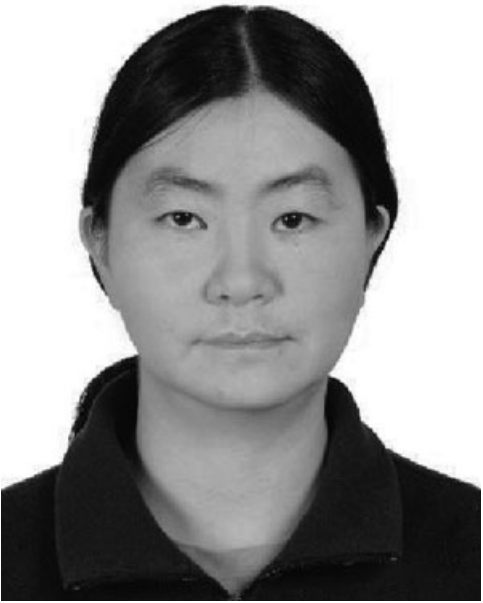}}]{Huawei Li}
(Senior Member, IEEE) received the B.S. degree from Xiangtan University in 1996 and the M.S. and Ph.D. degrees from the Institute of Computing Technology (ICT), Chinese Academy of Sciences (CAS), in 1999 and 2001, respectively. 

She has been a Professor at ICT, CAS since 2008, published over 180 papers, holds 27 Chinese patents, and her research focuses on VLSI/SOC testing, design verification, and design for reliability.
\end{IEEEbiography}

\end{document}